\documentclass[prl,aps,twocolumn,floatfix]{revtex4}
\usepackage{graphicx,graphics,psfrag,amsmath,calc,mathtools}
\usepackage{epsfig, bm}
\usepackage{color,comment}
\begin{document}

\title{Nematic-orbit coupling and nematic density 
waves in spin-1 condensates}

\author{Di Lao}
\email{dlao7@gatech.edu}
\author{Chandra Raman}
\author{C. A. R. S{\'a} de Melo}
\affiliation{
School of Physics, Georgia Institute of Technology, Atlanta, 
Georgia 30332, USA
}

\date{\today}

\begin{abstract}
We propose the creation of artificial nematic-orbit coupling in spin-1 Bose-Einstein condensates, in analogy to spin-orbit coupling. Using a suitably designed microwave chip, the quadratic Zeeman shift, normally uniform in space, can be made to be spatio-temporally varying, leading to a coupling between spatial and nematic degrees of freedom. A phase diagram is explored where three quantum phases with the nematic order emerge: easy-axis, easy-plane with single-well and easy-plane with double well structure in momentum space. By including spin-dependent and spin-independent interactions, we also obtain the low energy excitation spectra in these three phases. Lastly, we show that the nematic-orbit coupling leads to a periodic nematic density modulation in relation to the period $\lambda_T$ of the cosinusoidal quadratic Zeeman term. Our results point to the rich possibilities for manipulation of tensorial degrees of freedom in ultracold gases without requiring Raman lasers, and therefore, obviating light-scattering induced heating.
\end{abstract}
\maketitle

%
%

Ultracold atoms are a unique platform for exploring multi-faceted quantum magnetic behavior associated with spin. Some of the success stories in this arena include spinor BECs~\cite{ueda-2013}, where magnetic interactions play an important role, as well as systems with artificial spin-orbit coupling~\cite{spielman-2009, dalibard-2010, spielman-2011,sademelo-2011, pan-2014, ketterle-2017, zhai-2015, demarco-2015, zhang-2019,ye-2017,ye-2018, campbell-2016}, where independent-particle effects are primarily involved.  Yet a comprehensive experimental framework linking these two disparate regimes of spin physics in ultracold gases has been lacking.  In part, this is due to the fact that some of the richest behavior in spinor gases involves the dynamics of spin-nematic phases~\cite{ketterle-1998, machida-1998, zhou-2004, demler-2003,  affleck-2004, lett-2007, lett-2009, raman-2011, gerbier-2012, gerbier-2016, borgh-2014, symes-2017, kang-2019}.  These phases are special because they have a vanishing total magnetization vector $\langle \hat {\bf F} \rangle = 0$ and their order parameter is tensorial.  For a spin-1 system, the expectation value of the spin-quadrupole tensor operator ${\hat {\bf Q}}_{ij} = \frac{1}{2} \left( {\hat {\bf F}}_i {\hat {\bf F}}_j + {\hat {\bf F}}_j {\hat {\bf F}}_i \right) $ may act as an order parameter, where $i, j$ are the $\{x, y, z\}$ components of the spin-operator ${\hat {\bf F}}$~\cite{andreev-1984}.  Through interactions between atoms, such tensor objects naturally generate spin entanglement and strong correlations.  An important example of this is the reaction between two $|F=1,m = 0\rangle$ alkali atoms through $s$-wave scattering, that is $|1,0\rangle + |1,0\rangle \leftrightarrow |1,1\rangle + |1,-1\rangle$, which conserves $m_1 + m_2 = 0$ of atoms 1 and 2~\cite{sadler-2006,lucke-2011,gross-2011,bookjans-2011,vinit-2013,vinit-2018}. By contrast, the spin-orbit coupling achieved using Raman laser schemes does not readily lend itself to the study of pure spin-nematic objects, although a variety of other interacting many-body phases have been predicted~\cite{galitski-2008, ho-2011,  stringari-2012, baym-2012, stringari-2013,yamamoto-2017}.

\begin{figure} [tb]
$\begin{array}{l r}
{\rm (a)} \\  
\includegraphics[width=  0.8\columnwidth]{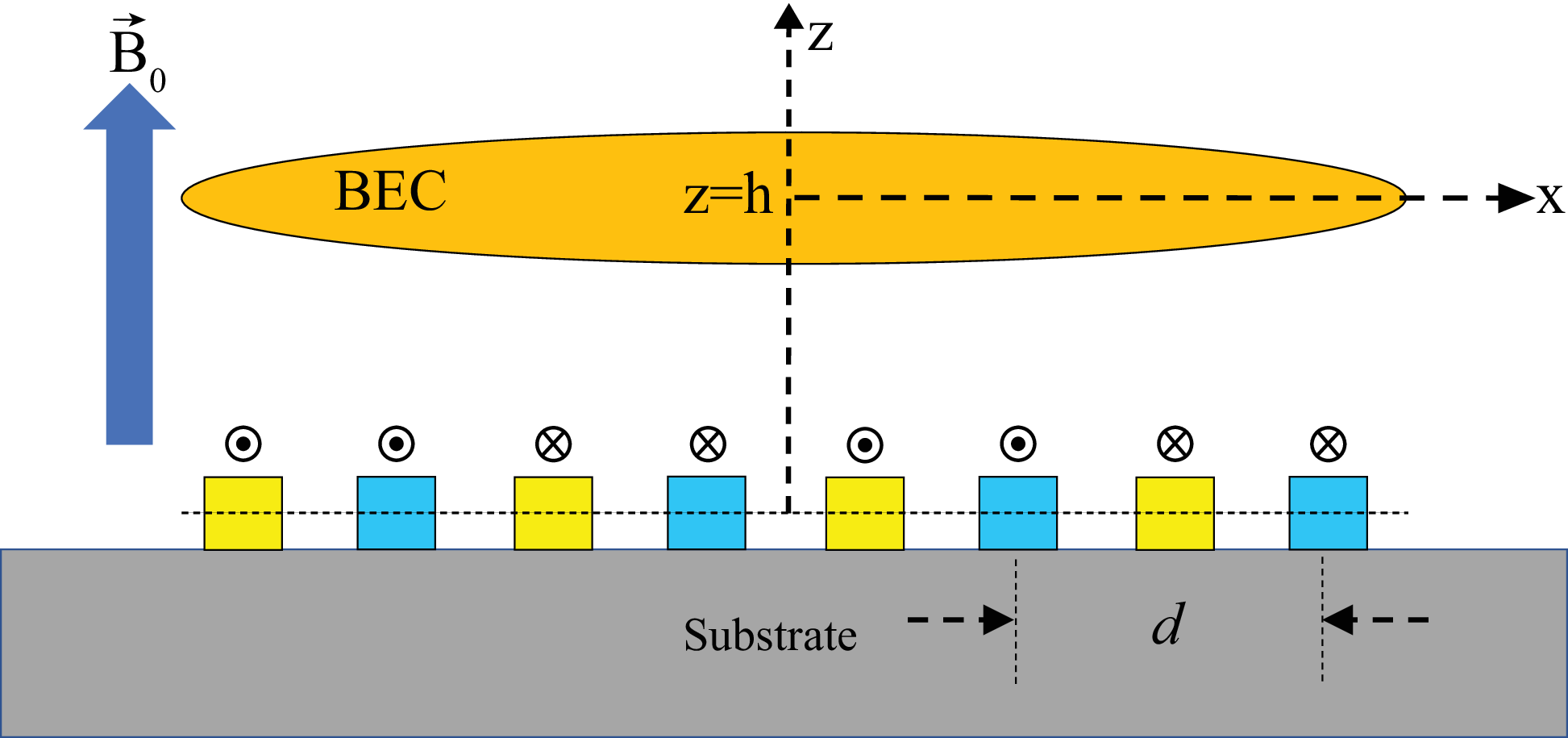} \\
{\rm (b)} \\    
\includegraphics[width=0.8 \columnwidth]{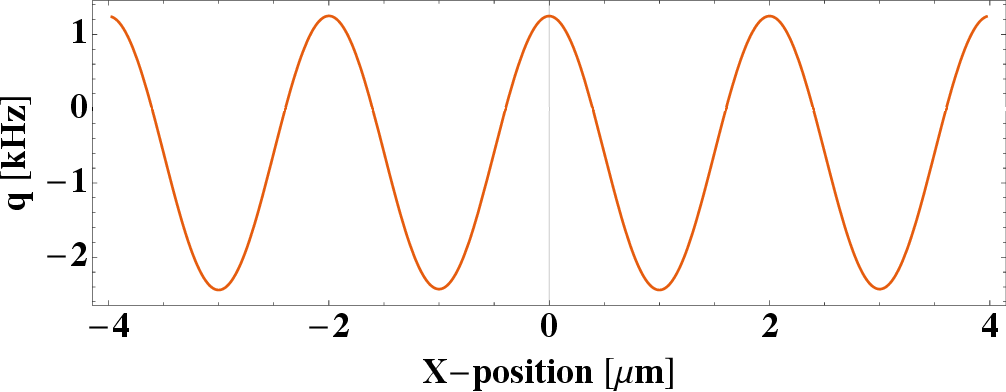}\\
\end{array}$
\caption{(Color Online).  Protocol for nematic-orbit coupling.  (a) Optically trapped Bose-Einstein condensate at a height $h$ above the centroid of a coplanar waveguide array (CPW).  The array is part of a monolithic microwave integrated circuit (MMIC) that modulates the quadratic Zeeman shift $q({\bf r},t)$ through the AC Zeeman effect.  Two interleaved sets of wires (yellow and blue) are energized with microwave currents whose amplitude is modulated in proportion to $\cos{\omega t}$ and $\sin{\omega t}$, respectively.  The result is a magnetic traveling wave creating a quadratic shift that varies nearly cosinusoidally as  $q + \Omega_c (z) \cos{(k_T x-\omega t})$.  $\omega$ is near resonance with the confinement along $z$, as discussed in \protect{\cite{supplementary-material}}. The spacing of each wire array is $d=2 \mu$m, a static field is $B_0 = 1.4$ Gauss and a microwave field amplitude of $B_1 = 0.1$ Gauss results from a current density amplitude per wire of $8.4 \times 10^4$ Amps/cm$^2$.  The microwave frequency is detuned by $\Delta = + 2$ MHz from the clock transition $|F = 1,m_F = 0\rangle \rightarrow |F = 2,m_F = 0\rangle$ at 1.77 GHz for $^{23}{\rm Na}$.
(b) Plot of $q(x,z=h,t=0)$ at $h = 2.5  \mu$m with $q = -600$ Hz, $\Omega_c (h) = 1840$ Hz, and $k_T = 2 \pi/(2 \mu {\rm m})$ . 
}
\label{fig:one}
\end{figure}

In contrast to spin-orbit coupling, in this work we explore nematic-orbit coupling, where the linear momentum of spin-1 bosonic atoms is coupled to the spin-nematic degrees of freedom.  Nematic spinor states have a zero expectation value for the spin vector $\langle {\hat {\bf F}} \rangle$ and nonzero quadrupole tensor $\langle {\hat {\bf Q}}_{ij} \rangle = \delta_{ij} - d_i d_j$, where ${\bf d}$ is the director.  Easy axis or easy plane states correspond to ${\bf d}$ aligned with either the $z$ direction or lying in the $xy$-plane, respectively.  Here, we propose an experimental setup to create nematic-orbit coupling between the center of mass of spin-1 bosons and the $zz$ component of the spin-quadrupolar operator ${\hat {\bf Q}}_{zz} = {\hat {\bf F}}_z^2$, as shown in Fig.~\ref{fig:one}.

In the setup shown in Fig.~\ref{fig:one}, a spatio-temporally varying quadratic Zeeman shift $q({\bf r},t){\hat {\bf F}}_z^2$ is created using a combination of a static bias field and a microwave field that is produced by a monolithic microwave integrated circuit (MMIC)~\cite{treutlein-2009}. 
After eliminating constant and linear terms in $\hat {\bf F}_z$ (see \cite{{supplementary-material}}), the effective independent particle Hamiltonian is
%
%
%
\begin{equation}
\label{eqn:independent-particle-hamiltonian}
{\hat H}_{{\rm IP}} 
= \int \mathrm{d} {\bf r} 
\sum_{a}
\psi_a^\dagger({\bf r})  
\bigg[ 
\frac {{\bf p}^2}{2m} {\hat {\bf 1}} 
+V({\bf r}) {\hat {\bf 1}} 
+
q({\bf r},t) {\hat {\bf F}}_z^2
\bigg] 
\psi_a ({\bf r}),
\end{equation}
where $\psi^\dagger_a({\bf r})$ is the creation operator of bosons at position ${\bf r}$ with spin components $a = \{\pm 1, 0\}$, ${\bf p}^2/2m$ is the kinetic energy, $V({\bf r})=V_{trap}(z)$ is the trap potential, $q({\bf r},t) = q + 2\Omega_c(z) \cos (k_T x - \omega t)$ is the resulting spatio-temporal modulation of the quadratic Zeeman shift with period $\lambda_T = 2\pi/k_T$ and $\hat {\bf 1}$ is the identity matrix. The modulation amplitude $\Omega_c(z) = \Omega_0 + \Omega_1 z$ defines the strength of the nematic-orbit coupling. Since $\Omega_c(z)$ varies linearly with the $z$-coordinate, it couples two discrete energy levels $\epsilon_{1},\epsilon_{2}$ with different parity, which are defined by the spin-independent trapping potential $V({\bf r})$. A resonance condition for the magnetic traveling wave can be achieved when $\omega \approx \omega_{12} \equiv (\epsilon_2-\epsilon_1)/\hbar$~\cite{supplementary-material}. Given the discrete nature of the spectrum along $z$, we write the field operators as $\psi_a({\bf r})=\sum_n \varphi_n (z) \psi_{n,a}(x,y)$, where $\varphi_n (z)$ is the eigenfunction of trap state $n=\{1,2\}$. Within the rotating wave approximation (RWA) and zero detuning $\omega-\omega_{12} = 0$, the Hamiltonian can then be rewritten in momentum space as (see \cite{supplementary-material}):
\begin{equation}
\label{eqn:independent-particle-hamiltonian-momentum-space}
{\hat H}_{\rm IP}
=
\sum_{{\bf k}_\perp n}
\hat \phi^\dagger_{{\bf k}_\perp n}
{\bf H}_{\rm D}
\hat \phi_{{\bf k}_\perp n}
+
\left [
\Omega
\hat \phi^\dagger_{{\bf k}_-,1}
{\hat {\bf F}}^2_z
\hat\phi_{{\bf k}_+,2}
+
{\rm H.c.} \right ]
.
\end{equation}
Here, 
$\hat \phi^\dagger_{{\bf k}_\perp n}=\left[ \phi_{n,1} ({\bf k}_\perp), \phi_{n,0} ({\bf k}_\perp), \phi_{n,\bar1} ({\bf k}_\perp)\right ]$ is the spinor creation operator with subscript $\bar1$ as a shorthand for $-1$, ${\bf k}_\perp=(k_x,k_y)$,
$
{\bf H}_D = 
\varepsilon_{\bf k} 
{\hat {\bf 1}} 
+ 
q {\hat {\bf F}}^2_z,
$
where
$
\varepsilon_{\bf k}
=
\hbar^2 k_\perp^2/(2m)
$ 
is the kinetic energy with $k_\perp = \vert {\bf k}_\perp \vert$, 
and ${\bf k}_\pm = {\bf k}_\perp \pm (k_T/2){\hat {\bf x}}$ 
are shifted momenta. The Hermitian conjugate (H.c.) term is 
$\Omega\hat \phi^\dagger_{{\bf k}_+,2} {\hat {\bf F}}^2_z
\hat \phi_{{\bf k}_-,1}$, where $\Omega = \int \mathrm{d}z \varphi^*_1(z)
[\Omega_1 z]\varphi_2(z)$ plays the role of a Rabi frequency (see~\cite{supplementary-material}). 
The diagonalization of Eq.\ (\ref{eqn:independent-particle-hamiltonian-momentum-space}) leads to a 
trivial eigenvalue $E_0= \hbar^2k_\perp^2/(2m)$
corresponding to spin component $ a = 0$, and to non-trivial 
eigenvalues
\begin{equation}
\label{eqn:eigenvalues-independent-particle}
E_{\alpha,\beta}({\bf k}_\perp)
=
q
+
\frac{\hbar^2}{2m} 
\bigg[ k_\perp^2
+
\frac{1}{4}k^2_T \bigg] 
\pm 
\sqrt{
\bigg[
\frac{\hbar^2}{2m}k_ x k_T
\bigg]^2
+
\Omega^2
}.
\end{equation}
The lower (higher) energy branch is labeled by $\alpha$ $(\beta)$, 
with corresponding eigenvectors 
\begin{equation}
\begin{pmatrix}
\chi_{a\alpha}({\bf k}_\perp) \\ \chi_{a\beta}({\bf k}_\perp)
\end{pmatrix}
=
\begin{pmatrix}
u_{+\alpha}({\bf k}_\perp) & u_{-\alpha}({\bf k}_\perp)\\ 
u_{+\beta}({\bf k}_\perp) & u_{-\beta}({\bf k}_\perp)
\end{pmatrix}
\begin{pmatrix}
\phi_{1,a}({\bf k}_-) 
\\ 
\phi_{2,a}({\bf k}_+)
\end{pmatrix},
\end{equation}
written as linear combinations of
$\phi_{1,a}({\bf k}_-) $
and 
$\phi_{2,a}({\bf k}_+) $.
Expressions for the coefficients $u_{\pm\alpha}({\bf k}_\perp)$ 
and $u_{\pm\beta}({\bf k}_\perp)$ are found 
in~\cite{supplementary-material}. The absolute minimum of all eigenvalues, where Bose-Einstein 
condensation occurs,  depends on parameters $q$ and $\Omega$, and is found in the lower band $\alpha$. We locate 
the minima of these energy bands by extremizing with respect to $k_x$. 
We work with dimensionless variables 
and set $k_T$ as the unit of momentum and $E_T = \hbar^2k^2_T/(2m)$ as
the unit of energy. The scaled parameters are $\widetilde q = q/E_T$, 
$\widetilde \Omega = \Omega/E_T$ and ${\widetilde {\bf k}}_\perp = {\bf k}_\perp/k_T$. 

\begin{figure} [tb]
\includegraphics[width = 0.8\columnwidth]{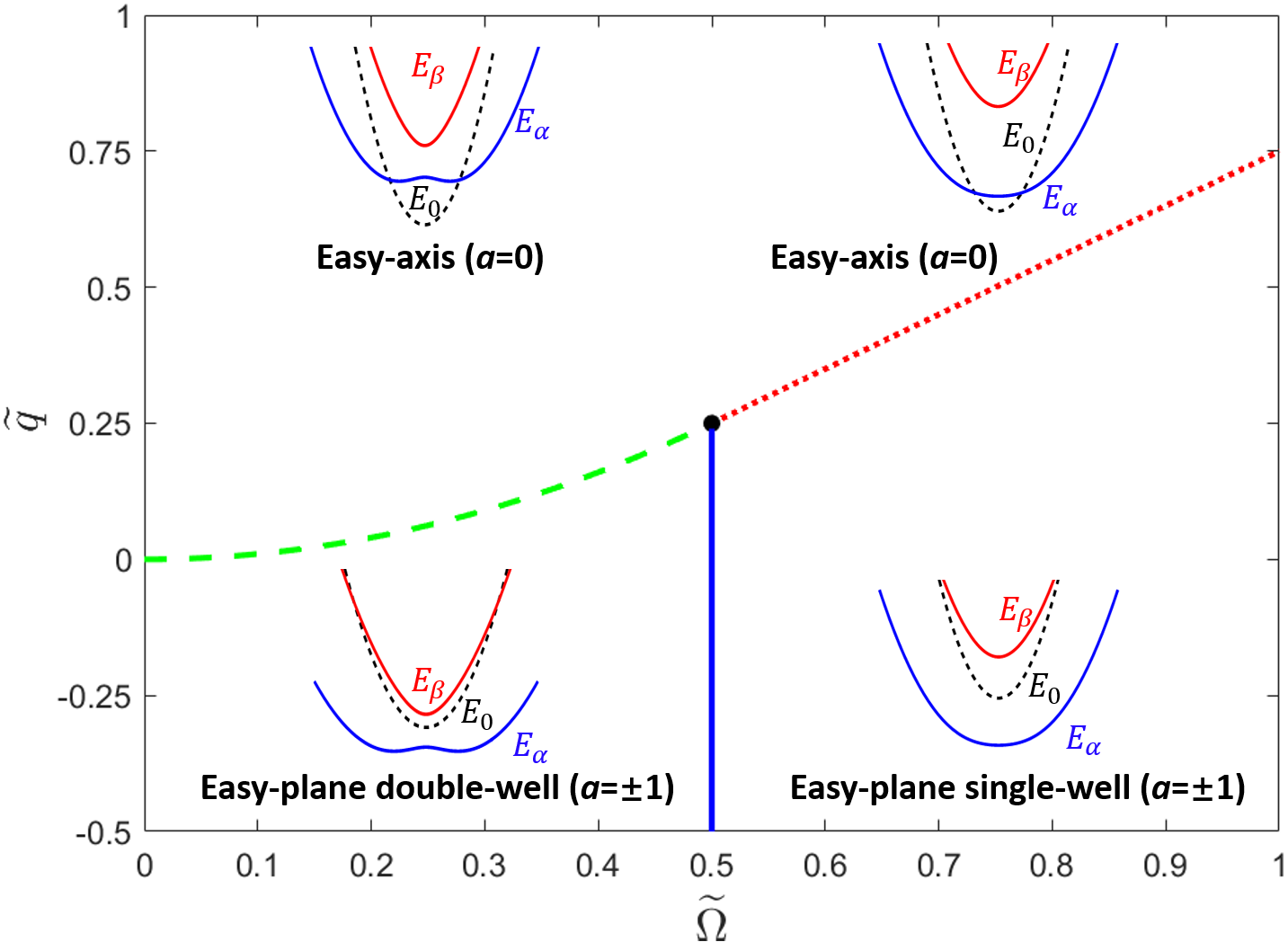}
\caption{(Color Online). 
Phase diagram of spin-1 Bose-Einstein condensates with nematic-orbit 
coupling. Shown are the ground state energies of Eq.~(\ref{eqn:eigenvalues-independent-particle}) in the ${\widetilde q}$ versus ${\widetilde \Omega}$ plane. 
The diagram is separated into three regions as discussed in the text. 
The modified band structures are shown at
four special coordinates 
$( {\widetilde q} = 0.75, {\widetilde \Omega} = 0.25 )$,
$( {\widetilde q} = 0.75, {\widetilde \Omega} = 0.75 )$ 
$( {\widetilde q} =-0.30, {\widetilde \Omega} = 0.25 )$, 
and 
$( {\widetilde q} =-0.30, {\widetilde \Omega} = 0.75$).}
\label{fig:two}
\end{figure}

In Fig.~\ref{fig:two}, we show the phase diagram of $\widetilde q$ versus
$\widetilde \Omega$ arising from Eq.~(\ref{eqn:eigenvalues-independent-particle}). 
The dashed-green line corresponds to the phase boundary 
${\widetilde q}_c ({\widetilde \Omega}) = {\widetilde \Omega}^2$
for ${\widetilde \Omega} < 1/2$, 
that separates an easy-axis nematic BEC 
at ${\widetilde {\bf k}_\perp} = {\bf 0}$ for spin component $a = 0$,
when ${\widetilde q} > {\widetilde q}_c ({\widetilde \Omega})$, 
from a double-well easy-plane nematic BEC for spin components $a = \pm 1$, 
when ${\widetilde q} < {\widetilde q}_c ({\widetilde \Omega})$. 
The dotted-red line describes the phase boundary 
${\widetilde q}_c ({\widetilde \Omega}) = {\widetilde \Omega} - 1/4$
for ${\widetilde \Omega} > 1/2$, 
that separates an easy-axis BEC at ${\widetilde {\bf k}_\perp} = {\bf 0}$ 
for spin component $a = 0$, when 
${\widetilde q} > {\widetilde q}_c ({\widetilde \Omega})$, 
from a single-well easy-plane nematic BECs for spin components $a = \pm 1$, 
when ${\widetilde q} < {\widetilde q}_c ({\widetilde \Omega})$. 
The solid-blue line ${\widetilde \Omega} = 1/2$ separates
the easy-plane nematic BECs in the $\alpha$ band into two sectors:
a) a double-well phase where condensation occurs at finite momenta
$({\widetilde k}_x,{\widetilde k}_y) = 
(\pm {\widetilde k}_0,0)$, with 
${\widetilde k}_0 = \sqrt{1/4-{\widetilde\Omega}^2}$, 
and 
b) a single-well phase where condensation occurs at zero momentum
$\widetilde {\bf k}_\perp = {\bf 0}$. 
The solid-black dot at coordinates 
$({\widetilde q}, {\widetilde \Omega}) = (1/4, 1/2)$ represents a triple point.

%
%
Next, we discuss the interaction Hamiltonian
$
\label{eqn:interaction-hamiltonian}
{\hat H}_{\rm int}  =
{\hat H}_0 + {\hat H}_2
$.
The first term is the spin-independent interaction ${\hat H}_0 = (c_0/2L^2_\perp){\widetilde H}_0$, with
\begin{equation}
\label{eqn:spin-independent-interaction-h0}
{\widetilde H}_0 = 
\sum_{\substack{{\bf k}_\perp {\bf k'}_\perp {\bf p}_\perp \\
a a^\prime \{n_i\}}}
C_{\{n_i\}} \Lambda_{n_1n_2}^{\dagger a a^\prime} ({\bf k}_{p-}, {\bf k}^\prime_{p+})
\Lambda_{n_3n_4}^{a^\prime a}  ({\bf k}^\prime_{p-}, {\bf k}_{p+}),
\end{equation}
where the subscripts $\{n_i\}$ denote the set of trapped states with quantum numbers $(n_1,n_2,n_3,n_4)$ that label the coefficients $C_{\{n_i\}} =\int \mathrm{d}z \varphi^*_{n_1}(z) \varphi^*_{n_2}(z)\varphi_{n_3}(z)\varphi_{n_4}(z)$. In Eq.~(\ref{eqn:spin-independent-interaction-h0}), the momenta are ${\bf k}_{p\pm} = {\bf k}_\perp \pm {\bf p}_\perp/2$ and ${\bf k}^\prime_{p\pm} = {\bf k}^\prime_\perp \pm {\bf p}_\perp/2$, and the operators are
\begin{equation}
\begin{split}
&\Lambda_{n_1n_2}^{\dagger a a^\prime} ({\bf k}_{p-}, {\bf k}^\prime_{p +})
=
\phi^\dagger_{n_1,a}({\bf k}_{p-}) \phi^\dagger_{n_2,a^\prime}({\bf k}^\prime_{p+}), \\
&\Lambda_{n_3n_4}^{a^\prime a}  ({\bf k}^\prime_{p-}, {\bf k}_{p +})
=
\phi_{n_3,a^\prime}({\bf k}^\prime_{p-}) \phi_{n_4,a}({\bf k}_{p+}).
\end{split}
\end{equation}

In the interaction Hamiltonain, the second term is the spin-dependent interaction ${\hat H}_2 = (c_2/2L^2_\perp){\widetilde H}_2$, with
\begin{equation}
\label{eqn:spin-dependent-interaction-h2}
{\widetilde H}_2 = 
\sum_{\substack{{\bf k}_\perp {\bf k}^\prime_\perp {\bf p}_\perp\\ aa^{\prime} b b^{\prime} \{ n_i \}}} 
C_{\{n_i\}} {\hat {\bf J}}^{ab}_{n_1n_4} ({\bf k}_{p-}, {\bf k}_{p+})
\cdot
{\hat {\bf J}}^{a^\prime b^\prime}_{n_2n_3} ({\bf k}^\prime_{p+}, {\bf k}^{\prime}_{p-})
\end{equation}
where the vector operators
\begin{equation}
\begin{split}
&{\hat {\bf J}}^{ab}_{n_1n_4}  ({\bf k}_{p-}, {\bf k}_{p+}) =
\phi^\dagger_{n_1,a} ({\bf k}_{p-}) 
{\hat {\bf F}}_{ab} \phi_{n_4,b} ({\bf k}_{p+}) \\
&{\hat {\bf J}}^{a^\prime b^\prime}_{n_2n_3} ({\bf k}^{\prime}_{p+}, {\bf k}^{\prime}_{p-})= 
\phi^\dagger_{n_2,a^\prime} ({\bf k}^{\prime}_{p+}) 
{\hat {\bf F}}_{a^\prime b^\prime} 
\phi_{n_3,b^\prime} ({\bf k}^{\prime}_{p-})
\end{split}
\end{equation}
contain the spin-1 matrices ${\hat {\bf F}}$.

The Hamiltonians ${\hat H}_{\rm IP} + {\hat H}_{\rm int}$ preserve the 
{\it magnetization} $m_z = n_{+1} - n_{-1}$, where $n_{\pm 1}$ is 
the density of bosons with spin component $a = \pm 1$, that is, $m_z$ 
is a conserved quantity of the total Hamiltoninan. 
From now on, we consider only $m_z = 0$, 
in which case a phase transition occurs at $\widetilde q_c = 0$ between 
the easy-plane nematic state $\vert \zeta_P \rangle$ 
$({\widetilde q} < {\widetilde q}_c)$ 
with spin-densities $n_0 = 0 $, $n_{+1} = n_{-1} \ne 0$, 
and the easy-axis nematic state $\vert \zeta_A \rangle$ 
$({\widetilde q} > {\widetilde q}_c)$ 
with spin-densities $n_0 \ne 0$, $n_{+1} = n_{-1} = 0$, as shown in 
Fig.~\ref{fig:two}, when ${\widetilde \Omega} = 0$~\cite{ketterle-1998, lett-2007,
raman-2011, ueda-2013, gerbier-2016}. 

The effects of nematic-orbit coupling are also present in 
the collective excitations.
First, we investigate the easy-axis nematic phase, 
where condensation occurs at $\widetilde {\bf k}_\perp ={\bf 0}$ 
for spin projection $a = 0$. The Bogoliubov excitation spectrum is then identical to a scalar condensate,
$
\varepsilon_{b}({\bf k}_\perp) =
\left[\varepsilon_{\bf k}
\left( 
\varepsilon_{\bf k}+ 2c_0 n_c
\right)
\right]^{1/2},
$
where $n_c$ is the total particle density and 
$\varepsilon_{\bf k} = \hbar^2 k_\perp^2/(2m)$ is
the kinetic energy.

Next, we consider the easy-plane nematic phase in the single-well regime when 
$\widetilde q \ll \widetilde \Omega -1/4$ and 
$\widetilde\Omega > 0.5$. We write the field operators 
$\phi_{n,a}$ in terms of $\chi_{a\alpha},\chi_{a\beta}$ as 
shown in~\cite{supplementary-material}.
Condensation occurs at ${\widetilde {\bf k}}_\perp = {\bf 0}$ for the $\alpha$-band only, thus 
we drop the $\alpha$ index from our notation. The resulting Bogoliubov Hamiltonian is
\begin{equation}
\label{eqn:single-well-easy-plane-Bogoliubov-Hamiltonian}
\hat H = 
G_{\rm sw}
+\frac{1}{2}\sum_{{\bf k}\neq 0} 
{\bf X}_{\bf k}^\dagger \\
\begin{pmatrix}
{\bf E}_{1} &  {\bf D} \\
{\bf  D}^\dagger & {\bf E}_{\bar 1}
\end{pmatrix}
{\bf X}_{\bf k}.
\end{equation}
The matrices for spin preserving processes are
\begin{equation}
{\bf E}_a
= 
\begin{pmatrix}
E_g({\bf k}_\perp) + c & fe^{i2\Phi_a} \\
fe^{-i2\Phi_a} & E_g({\bf k}_\perp) + c 
\end{pmatrix},
\end{equation}
where $a = \{+1, -1\}$ is represented by $\{1, {\bar 1}\}$,
$
E_g({\bf k}_\perp) 
= 
E_\alpha({\bf k}_\perp)
-
E_\alpha(0)
$ 
is a measure of the excitation energy of independent particles 
with respect to the minimum of the $\alpha$-band, 
$\Phi_a$ is the spin-dependent phase of the condensate 
in the $\alpha$-band at ${\bf k}_\perp = {\bf 0}$ 
and $c,f$ are proportional to the spin-preserving
interaction energy $(c_0 + c_2)n_c$.
The matrices for spin-flip processes are 
\begin{equation}
{\bf D}
= 
\begin{pmatrix}
de^{i(\Phi_1-\Phi_{\bar 1})} & ge^{i(\Phi_{\bar 1}+\Phi_1)} \\
ge^{-i(\Phi_1+\Phi_{\bar 1})} & de^{-i(\Phi_1-\Phi_{\bar 1})}\\
\end{pmatrix},
\end{equation}
and ${\bf D}^\dagger$, 
where $d$ and $g$ are proportional to the spin-flip
interaction energy $(c_0 - c_2)n_c$. 
Lastly, in Eq.~(\ref{eqn:single-well-easy-plane-Bogoliubov-Hamiltonian}),
$G_{\rm sw}$ is the ground state energy and 
$
{\bf X}^\dagger_{\bf k}
=
\begin{pmatrix}
\chi^\dagger_{1}({\bf k}_\perp) & 
\chi_{1}(-{\bf k}_\perp) & 
\chi^\dagger_{\bar 1}({\bf k}_\perp) & 
\chi_{\bar 1} (-{\bf k}_\perp)
\end{pmatrix}
$
is a vector operator, where 
$\chi^\dagger_a$ represents the creation operator in the $\alpha$-band.

\begin{figure} [tb]
\includegraphics[width = \columnwidth]{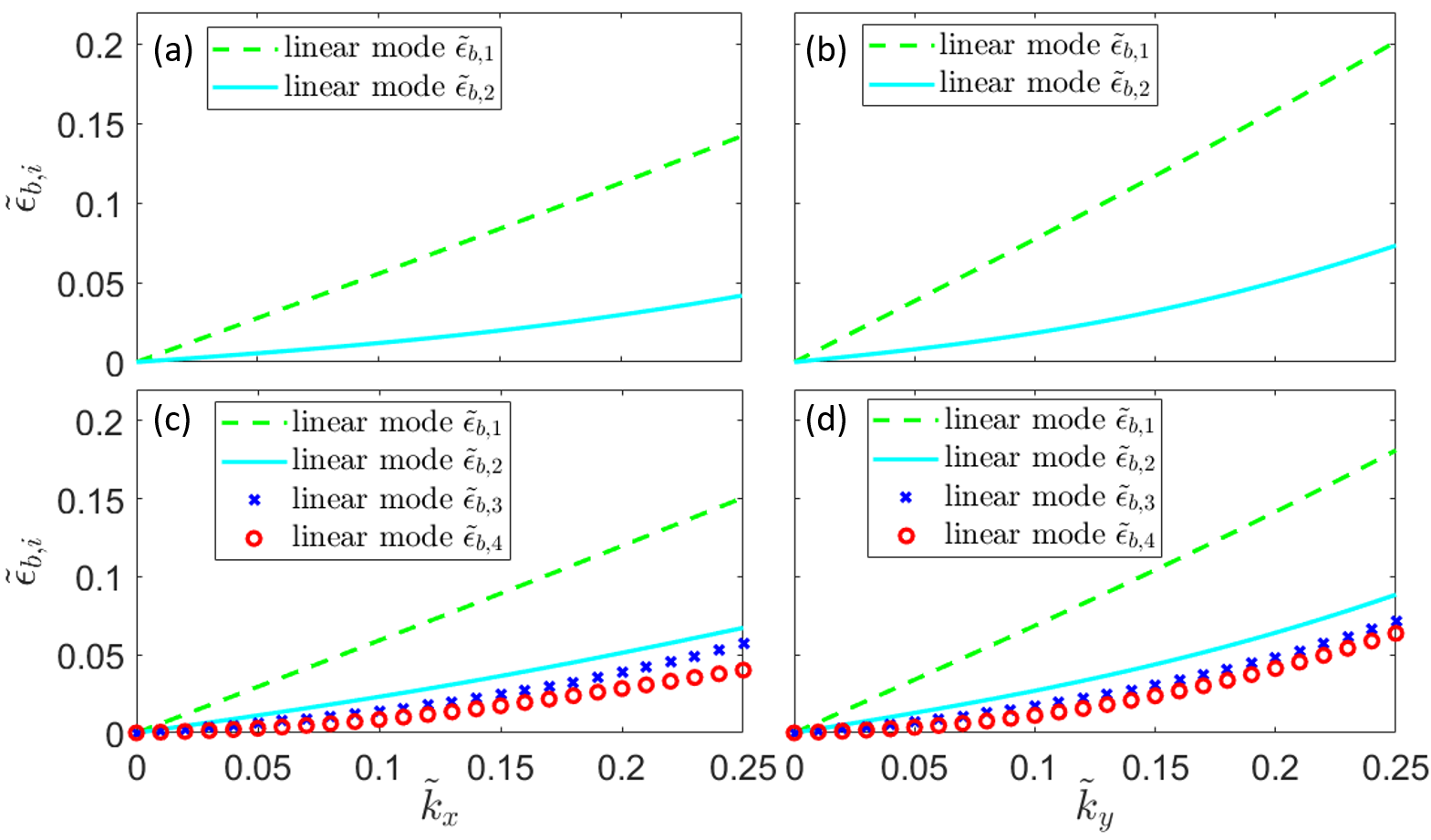}
\caption{(Color Online).  
Anisotropic collective excitation modes of a nematic-orbit coupled BEC.  Excitation energies ${\widetilde \epsilon}_{b, i} ({\bf k})$ 
for easy-plane nematic phases versus ${\widetilde k}_x$ and ${\widetilde k}_y$, 
with ${\widetilde k}_z = 0$ are shown in (a) and (b) for the single-well case
$(\widetilde q = -0.3, {\widetilde \Omega} = 1)$
and in (c) and (d) for the double well case
$(\widetilde q = -0.3, {\widetilde \Omega} = 1/4)$.
The other parameters are wavelength $\lambda_T = 2 \mu {\rm m}$, 
particle density $n_c = 2.5 \times 10^{13} {\rm cm}^{-3}$ and   
interaction constants $c_0 n_c/E_T = 0.168$ and 
$c_2 n_c/E_T = 6.74 \times 10^{-3}$. 
}
\label{fig:three}
\end{figure}

The positive eigenvalues in units of $E_T$ are
\begin{equation}
\label{eqn:easy-plane-bogoliubov-modes}
\begin{split}
&\widetilde \epsilon_{b,1} ({\bf k }_\perp) 
=\sqrt{\left[ \widetilde E_g({\bf k}_\perp)
+ (\widetilde c +\widetilde d) \right]^2-(\widetilde f + \widetilde g)^2
}, \\
&\widetilde \epsilon_{b,2} ({\bf k}_\perp)
=\sqrt{\left[\widetilde E_g({\bf k}_\perp)
+(\widetilde c - \widetilde d) \right]^2 - (\widetilde f - \widetilde g)^2
},
\end{split}
\end{equation}
where 
$
\widetilde E_g({\bf k}_\perp) = E_g ({\bf k}_\perp)/E_T
$
is a dimensionless independent particle energy,
$
\widetilde c 
=
(c_0 + c_2)n_c A_{\alpha} ({\bf k}_\perp) / (4E_T),
$ 
$
\widetilde f 
=
(c_0 + c_2)n_c B_{\alpha} ({\bf k}_\perp) / (4E_T)
$ 
are dimensionless spin-preserving interaction 
energies and 
$
\widetilde d 
= 
(c_0 - c_2) n_c A_{\alpha} ({\bf k}_\perp)/(4 E_T),
$
$
\widetilde g 
= 
(c_0 - c_2) n_c B_{\alpha} ({\bf k}_\perp)/(4 E_T),
$
are dimensionless spin-flip interaction energies.
Here, 
$
A_{\alpha} ({\bf k}_\perp) = 5/2 + 
\vert \widetilde \Omega \vert
/
\left[
 \sqrt{ {\widetilde k}_x^2 + {\widetilde \Omega}^2 }
\right]
$ 
and
$
B_{\alpha} ({\bf k}_\perp) = 2 + 
3\vert \widetilde \Omega \vert
/
\left[2
 \sqrt{ {\widetilde k}_x^2 + {\widetilde \Omega}^2 }
\right]
$ 
describe the anisotropic 
nature of the interactions induced by the
nematic-orbit coupling.
When ${\widetilde d} = {\widetilde g} = 0$, that is, $c_0 = c_2$, 
the matrix ${\bf D}$ of spin-flip processes vanishes and the spin-sectors
$\{ 1, {\bar 1} \}$ are uncoupled leading to two degenerate linear
modes at low momenta. Assuming that $c_0 > c_2 > 0$ like in $^{23}{\rm Na}$, 
we can understand a few limits 
from Eq.~(\ref{eqn:easy-plane-bogoliubov-modes}).
In the first mode, the sum 
${\widetilde c} + {\widetilde d}$ and 
${\widetilde f} + {\widetilde g}$
are proportional to the spin-independent interaction parameter $c_0$,
while in the second mode, the difference
${\widetilde c} - {\widetilde d}$ and  
${\widetilde f} - {\widetilde g}$
are proportional to the spin-dependent interaction parameter $c_2$.
Thus, the first mode is associated with density-density interactions
$c_0$, while the second is associated with spin-spin 
interactions $c_2$. 
We plot the excitation spectra
${\widetilde \epsilon}_{b,1} ({\bf k}_\perp) $ and 
${\widetilde \epsilon}_{b,2} ({\bf k}_\perp) $ versus $k_x$ in Fig.~\ref{fig:three}(a)
and versus $k_y$ in Fig.~\ref{fig:three}(b), 
with $c_0$ and $c_2$ values
for $^{23}{\rm Na}$~\cite{ho-1998}.

%
%

Lastly, we consider the easy-plane nematic phase in the double-well region,  
when $\widetilde q \ll {\widetilde \Omega}^2$ 
and $\widetilde\Omega < 0.5$. 
Condensation occurs in two degenerate minima at 
$\pm k_0 {{\bf \hat x}}$ of the $\alpha$-band.  
There are four excitation modes involving 
left $(L)$ and right $(R)$ wells and spin sectors
$a = \{ 1, {\bar 1} \}$. The Bogoliubov Hamiltonian becomes
\begin{equation}
\hat H = 
G_{\rm dw} 
+
\frac{1}{2}
\sum_{{\bf k}\neq 0} 
{\bf Y}_{\bf k}^\dagger 
\begin{pmatrix}
{\bf M}_{LL} & {\bf M}_{LR} \\
{\bf M}_{RL} & {\bf M}_{RR} 
\end{pmatrix}
{\bf Y}_{\bf k},
\end{equation}
where
$
{\bf Y}^\dagger_{\bf k}
=
\begin{pmatrix}
{\bf X}^\dagger_L ({\bf k}_\perp) &
{\bf X}^\dagger_R ({\bf k}_\perp)
\end{pmatrix}
$
is an eight-dimensional vector with 
four dimensional components 
$
{\bf X}^\dagger_{j} ({\bf k}_\perp)
=
\begin{pmatrix}
\chi^\dagger_{j 1}({\bf k}_\perp) & 
\chi_{j 1}(-{\bf k}_\perp) & 
\chi^\dagger_{j \bar 1}({\bf k}_\perp) & 
\chi_{j \bar 1} (-{\bf k}_\perp)
\end{pmatrix}
$
in the $j = \{ L, R \}$ sectors, and $G_{\rm dw}$ is the ground state energy.
The ${\bf M}_{ij}$ matrices are given 
in~\cite{supplementary-material} 
and the excitation spetrum is obtained numerically, 
but a qualitative understanding is possible. 
In each well there are equal numbers of 
atoms with spin components $a = \{1, {\bar 1}\}$,
that is, $n_{1L} = n_{1R}$ and $n_{{\bar 1}L} = n_{{\bar 1}R}$. 
When all interactions are present and all 
atoms oscillate in phase, this excitation corresponds to a 
center-of-mass motion with linear dispersion and lowest energy at low momenta, 
which is also anisotropic since the effective mass is heavier 
along $k_x$. When atoms with the same 
spin-projection $a$ oscillate in phase in both L and R wells, but 
out of phase with respect to their spin-projections, 
then a second linear mode arises with larger (larger) velocity
along $k_x$ $(k_y)$ in comparison to the center-of-mass mode.
When the spin-spin interactions are neglected and
atoms with spin-projection $a$ oscillate out of phase 
in $L$ and $R$ wells they produce two degenerate linearly 
dispersing modes. 
However, when spin-spin interactions are included the 
degeneracy of these modes is lifted producing a linearly 
dispersing mode with lower (higher) energy when the relative 
motion of $1$ and ${\bar 1}$ is in (out of) phase.
All four modes ${\widetilde \epsilon}_{b,1} ({\bf k}_\perp)$, 
${\widetilde \epsilon}_{b,2} ({\bf k}_\perp)$, 
${\widetilde \epsilon}_{b,3} ({\bf k}_\perp)$ and
${\widetilde \epsilon}_{b,4} ({\bf k}_\perp)$ of the excitation spectrum are shown 
in Fig.~\ref{fig:three}(c) and~\ref{fig:three}(d) for $^{23}{\rm Na}$ parameters.

%

%
%

Next, we analyze manifestations of the nematic-orbit coupling 
in real space and focus on the easy-plane nematic 
phases with $n_0 = 0$ and $n_{+1} = n_{-1} \ne 0$. 
Far below the phase boundary ${\widetilde q}_c ({\widetilde \Omega})$, 
the effective Hamiltonian is 
$
\hat H_{\rm EP}
= 
{\hat H}^\prime_{IP}
+
{\hat H}_{\rm I},
$
with
\begin{equation}
{\hat H}^\prime_{IP} =
\int \mathrm{d}^2 r_\perp
\begin{pmatrix}
{\hat \psi}^*_{1} & {\hat \psi}^*_{2}
\end{pmatrix}
\begin{pmatrix}
\frac {{\bf p}_\perp^2}{2m} + q\hat F^2_z  & \Omega e^{-i k_T x}\hat F^2_z \\
\Omega e^{i k_T x}\hat F^2_z & \frac {{\bf p}_\perp^2}{2m} + q\hat F^2_z 
\end{pmatrix}
\begin{pmatrix}
{\hat \psi}_{1}\\ {\hat \psi}_{2}
\end{pmatrix},
\end{equation}
where ${\hat \psi}^*_{n}=\left [\psi^*_{n,1}({\bf r}_\perp),\psi^*_{n,0}({\bf r}_\perp),\psi^*_{n,{\bar 1}}({\bf r}_\perp)\right ]$ represents the 2D condensate wavefunction in trap states with quantum number $n$.
The interaction Hamiltonian is ${\hat H}_{\rm I} = \int \mathrm{d}^3 r {\hat {\cal H}}_I$,
where  
\begin{equation}
{\hat {\cal H}}_{{\rm I}}
=
\frac{c_0}{2}
\left[ 
\vert {\bf \Psi}_1({\bf r})\vert^2 + \vert {\bf \Psi}_{\bar 1}({\bf r})\vert^2 
\right]^2 
+ 
\frac{c_2}{2}
\left[ 
\vert {\bf \Psi}_1({\bf r})\vert^2 - \vert {\bf \Psi}_{\bar 1}({\bf r})\vert^2 
\right]^2,
\end{equation}
with $c_0 > c_2 >0$ as in $^{23}{\rm Na}$, leading to the same local condensate densities, that is, 
$\vert {\bf \Psi}_{1}({\bf r}) \vert^2 = \vert {\bf \Psi}_{\bar 1}({\bf r}) \vert^2$.

\begin{figure}
\includegraphics[width = \columnwidth]{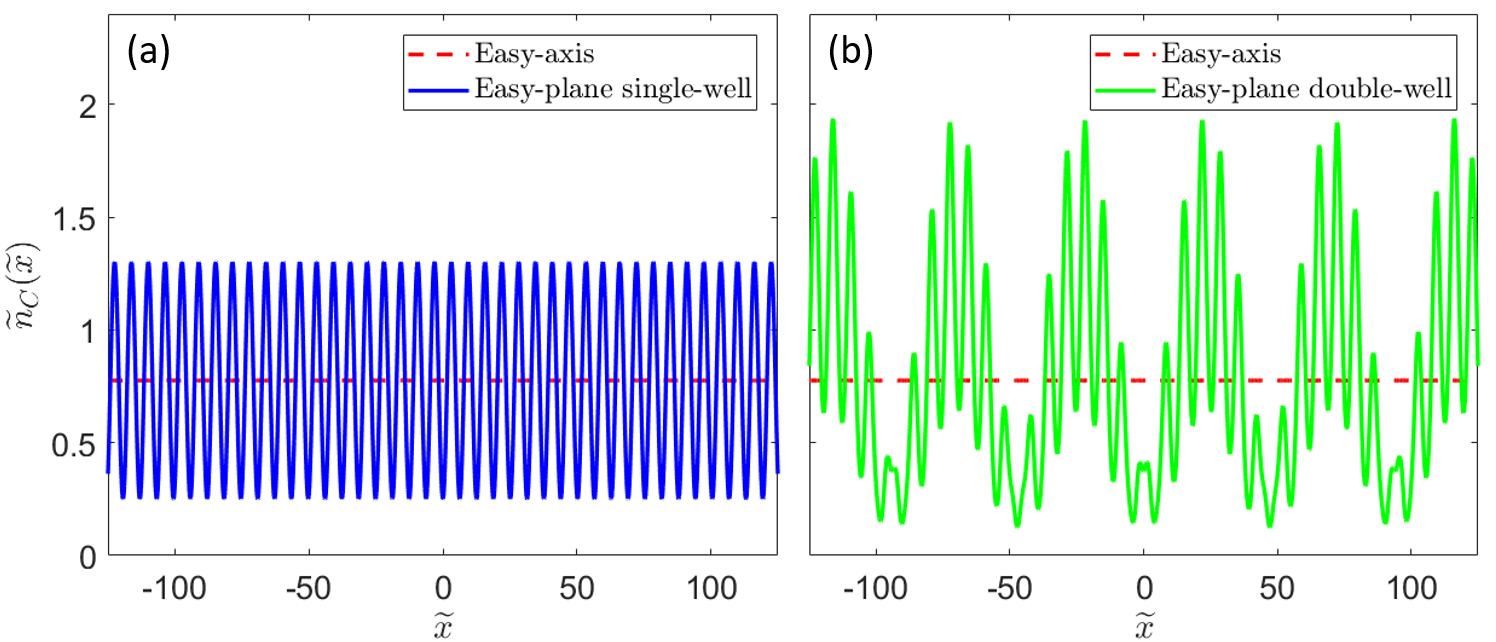}
\caption{(Color Online). 
Shown are easy-plane density modulations in real space for (a) single-well, with $\sigma = 0.7$, ${\widetilde z}=\pi/8$ and ${\widetilde \Omega} = 1$ (solid-blue line) and (b) double-well, with $\sigma = 0.7$, ${\widetilde z}=\pi/8$ and ${\widetilde \Omega} = 1/4$ (solid-green line).
The dashed-red line shows the uniform density profile 
of the easy-axis nematic phase.  The periodic modulation
in (a) is commensurate with $\lambda_T$, while in (b)
there are two periods, which are incommensurate with $\lambda_T$.
In (a) the period is $\lambda_T = 2 \mu m$, while in (b) 
the short period is $\lambda_+ = 2.14 \mu m$, while the long
period is $\lambda_- = 29.86 \mu m$. 
}
\label{fig:four}
\end{figure}

In the single-well phase, condensation occurs in the $\alpha$-band
at ${\widetilde {\bf k}}_\perp= {\bf 0}$. 
However, the wavefunction $\psi_a ({\bf r})$ in real space is a linear
combination of momentum shifted $(\pm (k_T/2) {\hat {\bf x}})$ condensates
with relative phase $\vartheta$~\cite{supplementary-material}, 
resulting in a spatial variation of the form
\begin{equation}
\label{eqn:easy-plane-single-well-wavefunction}
{\bf \Psi}_{a}({\bf r})
=
{\cal A}_{\rm sw}
e^{-i\frac{\vartheta}{2}}\bigg[e^{i\big(\frac{k_T}{2}x-\frac{\vartheta}{2}\big)}\varphi_2(z) - e^{-i\big(\frac{k_T}{2}x+\frac{\vartheta}{2}\big)}\varphi_1(z)\bigg],
\end{equation}
where $\varphi_{1,2}(z)$ are the trap states along $z$ direction and its period $\lambda_h = 2\pi/(k_T/2) = 2 \lambda_T$ commensurate to 
the period $\lambda_T$ of the periodic potential 
$q({\bf r},t)$. 
The phase $\vartheta = 0$~\cite{supplementary-material} is detemined by minimization of the free energy and ${\cal A}_{\rm sw}$ is obtained by normalizing the condensate density 
$
n_C ({\bf r}) 
= 
\sum_{a = \{1, {\bar 1} \}}
\vert 
\psi_a ({\bf r})
\vert^2
$ 
to the total number of condensed particles $N_C$~\cite{supplementary-material}.
Therefore, the dimensionless local condensate density ${\widetilde n}_C (\widetilde x)$ at some fixed $\widetilde z_0$, describing a easy-plane single-period nematic density wave (SPNDW), can be obtained by squaring the norm of Eq.~(\ref{eqn:easy-plane-single-well-wavefunction})~\cite{supplementary-material}.
%
%
${\widetilde n}_C ({\widetilde x})$  for $\sigma = 0.7$,
${\widetilde \Omega} = 1$ and $\widetilde z= \pi/8$ is plotted in Fig.~\ref{fig:four}(a), 
where $\widetilde x = k_T x$, $\widetilde z = (2\pi/L_z)z$ and $\sigma = N_C/N$ is the condensate fraction.
It is uniform apart from the periodic variation at the lattice period $\lambda_T$.

In the double-well phase, condensation occurs in the $\alpha$-band at 
${\widetilde {\bf k}}_\perp = \pm {\widetilde k}_0 {\hat {\bf x}}$. 
Thus, the wavefunction $\psi_a ({\bf r})$ 
in real space is a linear
combination of two single-well condensates with momenta 
$(k_0 \pm k_T/2){\hat {\bf x}}$ and 
phases $\vartheta$, $\vartheta_{LR}$~\cite{supplementary-material},
resulting in a spatial variation of the form
\begin{equation}
\label{eqn:easy-plane-double-well-wavefunction}
{\bf \Psi}_{a}({\bf r})
=
{\cal A}^\prime_{\rm dw}
\sum_{\substack{j = \pm \\ l=\pm}}
\bigg[
u_{j\alpha}(l{\widetilde k}_0)
e^
{i\big[(lk_0 + j \frac{k_T}{2})x 
-j\frac{\vartheta}{2}+l\frac{\vartheta_{LR}}{2}\big]
}\bigg]\varphi_j(z)
\end{equation}
with two periods $\lambda_{\pm} = 2\pi/\vert k_0 \pm k_T/2 \vert$, 
which are generically incommensurate with $\lambda_T$. Here, we denote ${\cal A}^\prime_{\rm dw}={\cal A}_{\rm dw}e^{-i\frac{\vartheta+\vartheta_{LR}}{2}}$, $\varphi_-(z)=\varphi_1(z)$ and $\varphi_+(z)=\varphi_2(z)$ for simplicity. The relative phase $\vartheta$, $\vartheta_{LR}$ were determined by minimizing the free energy numerically~\cite{supplementary-material}, resulting in $\vartheta=0$. The energy functional contains a rapid oscillation at the underlying period $\lambda_T$ as the system size $L_\perp$ is varied~\cite{supplementary-material}. We chose $k_T L_\perp = 250$ and $\vartheta_{LR} = 0$ to minimize the energy over this oscillation, with the results shown in Fig.~\ref{fig:four}(b). $\vartheta_{LR} = \pi$ achieved similar results for other $k_T L_\perp$. 
By squaring the wave function of Eq.~(\ref{eqn:easy-plane-double-well-wavefunction}), this leads to the dimensionless condensate density
%
describing a double-period nematic density wave (DPNDW) along $x$ direction shown 
in Fig.~\ref{fig:four}(b) for $\widetilde z = \pi /8$ (see~\cite{supplementary-material}). 

%
%

In conclusion, we have proposed a  mechanism for the creation of nematic-orbit coupling in spin-1 condensates and uncovered their phase diagram and excitation spectra. Our work connects orbital motion of atoms to the rich physics of spin-nematics, and opens up a new direction to explore strongly correlated spin-nematic states.  Future work may include higher spin systems and coupling to other tensor components ${\hat {\bf Q}}_{ij}$. Extension to higher dimensions could allow nontrivial topology to be explored, analogous to half-quantum vortices in ordinary nematics \cite{seo-2015}, which have parallels in solid state systems ~\cite{lagoudakis-2009, maeno-2011}.

\begin{acknowledgments}
This work was supported by NSF grant No. 1707654.  C. A. R. SdM acknowledges the support of the International Institute of Physics,
through its Visitor's Program.
\end{acknowledgments}


\pagebreak

\widetext
\begin{center}
\vskip 0.3cm
\textbf{ \large Supplementary Material}
\vskip 0.3cm 
\textbf{ \large
Orbit-nematic coupling in spin-1 condensate \\
}
\vskip 0.3cm
{
Di Lao$^*$, Chandra Raman and C. A. R. S{\'a} de Melo \\
{\it School of Physics, Georgia Institute of Technology, Atlanta, 
Georgia 30332, USA}
}
\end{center}

\vskip 0.3cm

In this supplementary material we provide additional 
information for each call made in the main text as 
reference~[41].
Thus, first, we discuss how to generate the nematic-orbit coupling using a 
chip design to produce a spatially dependent quadratic Zeeman shift.
Second, we investigate the independent particle Hamiltonian and third, 
the eigenvectors of the Hamiltonian are computed.
Fourth, we analyze the Bogoliubov spectrum of the  
easy-plane nematic phase in the single-well regime. 
Fifth, we perform a similar discussion for the easy-plane nematic phase 
in the double-well regime. Sixth, we investigate the Bogoliubov
spectrum for the easy-plane nematic phase in the double-well case.
And, lastly, we provide a real space description of the
easy-plane nematic phases in the single-well and double-well regimes. 

\begin{center}
\vskip 0.3cm
\textbf{Spatially varying quadratic Zeeman shift}
\vskip 0.3cm
\end{center}

Artificial nematic-orbit effects require the coupling of spatial coordinates to the 
spin quadrupole tensor. The simplest type of nematic-orbit coupling can be 
achieved via the production of a spatially varying quadratic Zeeman shift.
Thus, we describe a possible experimental implementation of a spatially varying 
quadratic shift via a setup that is similar to the experiment described 
in Ref.~\cite{treutlein-2009}, whose theoretical description and notation 
we follow here.  We consider an alkali atom at a fixed location $x,z$ and $y = 0$ 
that interacts with the local magnetic field.  The latter is the sum of a uniform 
static bias field $B_b \hat{{\bf z}}$ and a microwave field ${\bf B}_1(x,z,t)$, see Fig.~1 in the main text.  
The resulting atomic Hamiltonian ${\hat H_{\rm A}}$ depends on the total electronic 
(${\hat {\bf J}} = {\hat {\bf L}}+{\hat {\bf S}}$) and nuclear (${\hat {\bf I}}$) angular momenta.  
These combine to form the total angular momentum ${\hat {\bf F}} = {\hat {\bf J}} + {\hat {\bf I}}$.  
In the electronic ground state ${\bf L}={\bf 0}$ and hence 
\[ {\hat H_{\rm A}} = A {\hat {\bf I}}\cdot {\hat {\bf S}} 
+ 
\frac{g\mu_B}{\hbar} {\hat {\bf S}} \cdot 
\left[ B_0 \hat{\bf z}
+
{\bf B}_1(x,z,t)
\right], \]
where $A$ is the hyperfine coupling constant, $g \approx 2$ is the electron g-factor, 
and $\mu_B = e \hbar/(2 m_e)$ is the Bohr magneton.  For atoms such as 
$^{23}{\rm Na}$ or $^{87}{\rm Rb}$ the nuclear spin $I = 3/2$, which leads to lower and 
upper hyperfine levels $F = 1$ and $F = 2$, respectively.  
Their energy splitting is $W_{\rm hf} = 2A$. In the analysis above, we assume that 
the hyperfine energy splitting is much larger than the Zeeman interaction with the 
external bias field ($W_{\rm hf} \gg \mu_B B_0 $) so that we may use the good quantum numbers $F$ and $m_F$.  
Moreover, we also assume that $B_0 \gg B_1$ so that the linear Zeeman energy far 
exceeds the quadratic Zeeman energy shifts, as well as, those due to the 
microwave field.  For example, in the case of $^{23}$Na, 
$W_{\rm hf} = \hbar  \omega_0 = h \times  1.77{\rm GHz}$ while at a bias 
field of $B_0 = 1.4{\rm G}$, the linear Zeeman shift is 
$\mu_B B_0/(2 h) = 1{\rm MHz}$.  The latter is then larger than both the $\sim$ kHz quadratic shifts induced 
by the microwaves as well as the atomic trapping frequency along the tight $z$-direction, which is in the range of 10-100 kHz.  We have also neglected the 
interaction ${\hat{\bf I}}\cdot {\bf B}$ between the nuclear spin ${\hat{\bf I}}$ and
the magnetic field ${\bf B}$, which is in the 1 Hz range.  


For a near resonant microwave field ${\bf B}_1(x,z,t) = {\bf B}_1(x,z) \cos{\omega_c t}$, 
we make the rotating wave approximation whereby 
$|\Delta_0| \ll  \omega_c +\omega_0$, where $\Delta_0 = \omega_c -\omega_0$ 
is the detuning from the clock transition between states 
$\vert F, m_F\rangle = \vert 1,0\rangle$ and $\vert 2,0\rangle$.  
Using the basis $\vert F, m_F\rangle$ with quantization axis taken to be along the 
direction $\hat{\bf z}$ of the bias field, we express the atomic Hamiltonian as 
\begin{equation}
H_{\rm A} 
= 
\sum_{m_1}  (\hbar \omega_L m_1 +\frac{1}{2} \hbar \Delta_0) 
\vert 1,m_1\rangle \langle 1,m_1 \vert
+
\sum_{m_2} (\hbar \omega_L m_2 -\frac{1}{2} \hbar \Delta_0) 
\vert 2,m_2\rangle \langle 2,m_2 \vert  
+ 
\sum_{m_1,m_2} 
\left[ 
\frac{1}{2}\hbar \Omega_{1,m_1}^{2,m_2} 
\vert 2,m_2\rangle \langle 1,m_1 \vert + {\rm h.c.} 
\right] 
\end{equation}
where the Larmor frequency $\omega_L = \mu_B B_0/(2 \hbar)$ is associated with the linear Zeeman term.  The microwave field couples together states 
$\vert F = 1,m_1\rangle$ to $\vert F = 2,m_2\rangle$ with Rabi frequency
\[ 
\Omega_{1,m_1}^{2,m_2} 
= 
\frac{2 \mu_B}{\hbar} \langle 2,m_2\vert {\bf B}_1\cdot {\hat{\bf S}} \vert 1,m_1\rangle 
\]
and detuning 
\[ 
\Delta_{1,m_1}^{2,m_2} = \Delta_0 +(m_1-m_2)\omega_L. 
\]
We picture, using coordinates defined in Fig.\ 1 of the main text, the microwave field from a periodic array of alternating current wires arranged 
along the $x$-direction with spacing $d$, such that the base period of the current is $2d$.  $z$ is the coordinate perpendicular to the surface, which is parallel to the applied field ${\bf B}_0 = B_0 \hat{\bf z}$ that defines the quantization axis.  Due to the difference in Clebsch-Gordan coefficients for microwave fields parallel to and perpendicular to ${\bf B}_0$, a periodically varying magnetic field orientation results in a cosinusoidally varying quadratic Zeeman shift $\propto \cos (k_T x-\omega t)$, as we detail below.  

In general a full electromagnetic calculation would be needed to determine the microwave field pattern, however, the quasistatic approximation may be used when the wire spacing and lengths $d,L \leq 0.1{\rm cm}$ are much smaller than the microwave wavelength corresponding to the hyperfine splitting:  $\lambda \simeq 17{\rm cm}$ at $\omega_c = 2 \pi \times 1.77$GHz.  In this case, the magnetic field amplitude may be written in terms of a scalar potential ${\bf B}_1 = -\nabla \Phi_M$ satisfying Laplace's Equation $\nabla^2 \Phi_M = 0$.  The exact solution is a sum $\Phi_M = \sum_{l=1}^{\infty} \Phi_l \cos({l \pi x/d})\exp({-l \pi z/d})$ over all spatial harmonics of the base frequency $\pi/d$.  The resulting quasi-static magnetic field for $z>0$ is
\[
{\bf B}_1(x,z,t) = \sum_{l=1}^{\infty} B_l e^{-l\pi z/d} 
\left(
\sin{\left ( \frac{l \pi x}{d}\right ) } \hat{\bf x} -\cos{\left ( \frac{l \pi x}{d}\right )} \hat{\bf z} 
\right) \cos{\omega_c t} 
\]
For distances $z\sim d$ or greater above the wires, only the first harmonic $l = 1$ survives, resulting in
\begin{equation}
{\bf B}_1(x,z,t) \approx B_1 e^{-\pi z/d} \left(
\sin{\left ( \frac{\pi x}{d}\right ) } \hat{\bf x} -\cos{\left ( \frac{\pi x}{d}\right )} \hat{\bf z} 
\right) \cos{\omega_c t} 
\label{eq:field}
\end{equation}
where $B_1$ is proportional to the applied currents.  While the base spatial frequency is $\pi/d$, the quadratic shift varies as the square of $B$, resulting in a spatial frequency $k_T \equiv 2\pi/d$ that is twice as large, as detailed below.  

To achieve the required magnetic traveling wave with quadratic shift $\propto \cos (k_T x-\omega t)$, we utilize a second wire array that is shifted by $x = d/2$ with respect to the first one, as shown in Fig.\ 1 of the main text.  Performing a low frequency modulation of the microwave currents $I_1$ and $I_2$ in time as $I_1(t) = I_0 \cos({\omega t})$ and $I_2(t) = I_0 \sin({\omega t})$, we achieve a magnetic wave that travels along the $x$-direction.  Thus we can replace the coordinate $x$ by $x-v t$ where the velocity $v = \omega d/\pi = 2 \omega/k_T$.  Using a coordinate $z^\prime = z-h$ relative to the trapping point at $z = h$, the field experienced by the atoms for tight $z$-confinement is approximately
\begin{equation}
{\bf B}_1(x,z,t) = B_1 e^{-\pi h/d}\left (1-\frac{\pi z^\prime}{d} \right )\left[
\sin{\left ( \frac{\pi (x-vt)}{d}\right ) } \hat{\bf x} -\cos{\left ( \frac{\pi (x-vt)}{d}\right )} \hat{\bf z} 
\right] \cos{\omega_c t} 
\label{eq:field2}
\end{equation}
Numerical calculations using arrays of 201 wires confirm that for $z \geq d$ Eqns.\ (\ref{eq:field}) and (\ref{eq:field2}) are correct at the 1 \% level or better.  
Depending on the parameters in the phase diagram of Fig.~2 of the main text, one will also need an auxiliary {\em uniform} microwave field ${\bf B}_1^\prime \cos{\omega_c^\prime t}$ that creates a uniform quadratic shift $q_{M0}$ to adjust the offset $q$ that appears on the vertical axis.  This field could be applied from the top of the structure.

Now that the field ${\bf B}_1$ has been defined, we can diagonalize the Hamiltonian assuming that the microwave fields are small perturbations to the DC field ${\bf B}_0$.  In this limit the energy eigenvalues are labeled by the quantum numbers $F, m$ with 
$m = \{1, 0, \bar 1 \}$, but are dressed by the local magnetic field, resulting 
in adiabatic eigenstates.  For these eigenstates the energy of state $F=1,m$ is given by~\cite{bloch-2006}
\[ 
E_m (x,z) 
= 
\sum_{m_2} \frac{\hbar |\Omega_{1,m}^{2,m_2} (x,z)|^2}{4 \Delta_{1,m}^{2,m_2}}, 
\]
where as defined earlier, the $\Omega$ are Rabi frequencies proportional to the square of the field components perpendicular to and parallel to the applied field, ${\bf B}_1\cdot \hat{x}$ and ${\bf B}_1\cdot \hat{z}$, respectively.  The constants of proportionality are the squares of Clebsch-Gordan coefficients, as detailed in \cite{bloch-2006}.  When written in matrix form the above expression becomes
\begin{equation}
\begin{pmatrix}
E_1 (x, z) & 0 & 0 \\
0 & E_0 (x, z) & 0 \\
0 & 0 & E_{\bar 1} (x, z)
\end{pmatrix}
= 
\alpha (x, z){\hat {\bf 1}} 
+ 
\beta (x,z) {\hat {\bf F}}_z
+
q_M (x,z) {\hat {\bf F}}_z^2,
\end{equation}
where $\alpha (x, z) = E_0 (x, z)$ is a spatially varying state independent shift,
$\beta (x, z) = \left[ E_{1} (x,z) - E_{\bar 1} (x, z) \right]$ is a spatially 
varying linear shift, and 
\[ 
q_M(x,z) = \frac{1}{2} \left[ E_1(x,z)+E_{\bar 1}(x,z)- 2 E_0(x,z) \right]
\]
is the spatially varying quadratic shift due to the microwaves.  We isolate this term using the following procedure.  The state independent term $\alpha$ can be removed by superimposing a $\simeq 1$ kHz depth far-detuned optical lattice to the optical trapping potential whose depth is typically $20$ kHz for sodium atoms.  The linear shift is around 500 Hz, of the same order, and can easily be removed by adding a tiny, spatially varying 0.3 mG offset to the static field $B_0 = 1.4$ Gauss.  DC currents co-propagating with the microwave currents in the CPW can achieve this.  With this cancellation, only the desired nematic-orbit coupling $q_M (x, z) {\bf F}_z^2$ remains. To this, we add the quadratic shift due to the uniform bias field,
\[
q_{DC} 
= 
\frac{(g \mu_B)^2}{\Delta W (1+2I)^2} \times B_0^2 \approx {\rm 277 Hz/G}^2 \times B_0^2 
\]
and the spatially independent microwave field $q_{M0}$.  Combining all the equations above yields the final expression for the total quadratic shift.  Redefining the $z$ coordinate about the trap center, $z \equiv z^\prime$, we get 
\begin{equation}
q(x,z,t) = q + 2 \Omega_c (z) \cos(k_T x - \omega t)
\end{equation}
where $\Omega_c (z) = \Omega_0 + \Omega_1 z$ is defined as one-half of the amplitude of the cosinusoidal spatial variation of $q_M (x, z = h)$ with period $\lambda_T = 2\pi/ k_T$, and $\Omega_1 = - 2 \Omega_0 \pi/d$.  The constant factor is $q = q_{DC} + q_{M0}$.  This is the final expression for the nematic-orbit coupling used throughout the main text.

\begin{center}
\vskip 0.3cm
\textbf{Independent particle Hamiltonian}
\vskip 0.3cm
\end{center}

In our system, we have the following independent particle Hamitonian
\begin{equation}
\hat H =\frac{{\bf p}^2}{2m} {\hat {\bf 1}} +V_{trap}(z) {\hat {\bf 1}} +\bigg[q+2\Omega_c (z) \cos (k_T x-\omega t)\bigg]\hat F_z^2,
\end{equation}
where $\hat F_z$ is the $z$ component of spin-1 operator and $\hat{\bf 1}$ is the identity matrix.  We envision a box trapping potential $V_{trap}(z)$, which together with the linear variation of $\Omega_c (z)$ shown above, results in a resonance condition for the magnetic traveling wave between lowest levels $\epsilon_{1,2}$ of opposite parity.   Alternately, $V_{trap}$ could represent a mostly harmonic confinement from an optical lattice in which a single site has been isolated, with a small anharmonicity that isolates two levels.  To illustrate the basic feasibility without too much experimental detail, a box potential with width $l_{box} = 250$ nm at $h = 2.5$ $\mu$m above the wire array would result in energies $\epsilon_n = n^2 \times \frac{\pi^2 \hbar^2}{2Ml_{box}^2}$ and resonant frequencies of $\omega_{12} = (\epsilon_2-\epsilon_1)/\hbar \approx 2\pi \times 140$kHz.  This is smaller than the $1$ MHz Larmor precession frequency so that spin-transitions do not occur.  However, it is much larger than the energy level variations of the motional states due to the quadratic Zeeman effect (kHz), so that they remain adiabatic.  This allows us to apply a rotating wave approximation to the two coupled levels $\epsilon_{1,2}$ in which $\omega - \omega_{12} \ll \omega_{12}$ that allows us to average over the fast variations at frequency $\omega$ as we show below.  

Introducing field operators $\hat \psi^\dagger ({\bf r})=( \psi^\dagger_1 ({\bf r}), \psi^\dagger_0 ({\bf r}), \psi^\dagger_{\bar1} ({\bf r}))$ and applying second quantization,
\begin{equation}
\hat H =  \int \mathrm{d}^3 r \bigg[\hat \psi^\dagger({\bf r}) \frac {{\bf p}^2}{2m} {\hat {\bf 1}}  \hat\psi ({\bf r})+\hat\psi^\dagger({\bf r})V_{trap}(z){\hat {\bf 1}}  \hat\psi^\dagger({\bf r}) + \hat\psi^\dagger ({\bf r}) \bigg[q+2\Omega_c (z) \cos (k_T x-\omega t)\bigg]\hat F_z^2 \hat\psi ({\bf r})\bigg]
\end{equation}
We separate the $(x,y)$ and $z$ coordinates in the annihilation and creation operators as
\begin{equation}
\psi_a({\bf r})=\sum_n \varphi_n (z) \psi_{n,a}({\bf r}_\perp),
\end{equation}
where ${\bf r}_\perp=(x,y)$ and $n$ denotes the two trapped states with the lowest energy, therefore, $n=1,2$.  With $\Omega_c (z) = \Omega_0 + \Omega_1 z$ as defined in the previous section, the Hamiltonian becomes
\begin{equation}
\begin{split}
\hat H &= \sum_{n} \int \mathrm{d}^2 r_\perp \bigg[\hat \psi_n^\dagger({\bf r}_\perp) \frac {{\bf p}_\perp^2}{2m} {\hat {\bf 1}}  \hat \psi_n({\bf r}_\perp) +\hat\psi^\dagger_n({\bf r}_\perp) (\varepsilon_n{\hat {\bf 1}}  + (q + 2 \Omega_0 \cos (k_T x-\omega t)) \hat{F}_z^2) \hat\psi_n({\bf r}_\perp) \bigg]\\
&+ \sum_{n\neq n'} \int \mathrm{d}^2 r_\perp \hat \psi^\dagger_n({\bf r}_\perp) 2\Omega (1-\delta_{nn'})\cos (k_T x-\omega t)\hat F_z^2\hat \psi_{n'}({\bf r}_\perp),
\end{split}
\end{equation}
where $\hat \psi^\dagger_n ({\bf r}_\perp)=( \psi^\dagger_{n,1} ({\bf r}_\perp), \psi^\dagger_{n,0} ({\bf r}_\perp), \psi^\dagger_{n,\bar1} ({\bf r}_\perp))$ and $\int \mathrm{d}z \varphi_{n}(z)[\Omega_1 z] \varphi_{n'}(z)\equiv \Omega (1-\delta_{nn'})$ since $\Omega_c (z)$ depends linearly on $z$ and the two states $n=1,2$ have different parity.  $\varepsilon_n$ are the eigenenergies of the two trapped states, and ${\bf p}^2_\perp/(2m)= p_x^2/(2m) + p_y^2/(2m)$. Then we perform the unitary transformation $\hat U=e^{i\omega t|2\rangle \langle2|}$ to this Hamiltonian and apply the rotating wave approximation to $\cos{(k_T x -\omega t)}$.  In the latter, we can eliminate all the fast terms oscillating at $\omega,2\omega$, which do not survive the temporal average over the fast timescale $\sim 2\pi/\omega$. These include both the counter-rotating terms as well as term containing $\Omega_0$.  Subtracting a constant energy $(\varepsilon_1+\varepsilon_2-\omega)/2$ and defining the detuning $\delta = \omega -\omega_{12}$, the result is a $2 \times 2$ matrix in the basis of $z$-confinement:
\begin{equation}
\begin{split}
\hat H_{r} =\int \mathrm{d}^2 r_\perp
\begin{pmatrix}
\hat \psi_1^\dagger({\bf r}_\perp) & \hat \psi_2^\dagger({\bf r}_\perp)
\end{pmatrix}
\begin{pmatrix}
\frac {{\bf p}_\perp^2}{2m} + q\hat F^2_z + \frac{\hbar\delta}{2} & \Omega e^{-i k_T x}\hat F^2_z \\
\Omega e^{i k_T x}\hat F^2_z & \frac {{\bf p}_\perp^2}{2m} + q\hat F^2_z -\frac{\hbar\delta}{2}
\end{pmatrix}
\begin{pmatrix}
\hat \psi_1({\bf r}_\perp) \\ \hat \psi_2({\bf r}_\perp)
\end{pmatrix},
\end{split}
\end{equation}
Then we transform the Hamiltonian into momentum space by introducing field operators in momentum space
\begin{equation}
\psi_{n,a} ({\bf r}_\perp)=\frac{1}{L_\perp}\sum_{{\bf k}_\perp }\phi_{n,a}({\bf k}_\perp)e^{i{\bf k}_\perp\cdot{\bf r}_\perp},
\end{equation}
where ${\bf k}_\perp = (k_x,k_y)$.  Shifting the momentum ${\bf k}_\perp$ by $(\pm k_T/2)\hat x$, the Hamiltonian can be written as
\begin{equation}
\begin{split}
\hat H_{r} =\sum_{{\bf k}_\perp}
\begin{pmatrix}
\hat\phi^\dagger_1({\bf k}_-) & \hat\phi^\dagger_2({\bf k}_+)
\end{pmatrix}
\begin{pmatrix}
\frac {\hbar^2({\bf k}_\perp-(k_T/2)\hat x)^2}{2m} + q\hat F^2_z + \frac{\hbar\delta}{2} & \Omega \hat F^2_z \\
\Omega \hat F^2_z & \frac {\hbar^2({\bf k}_\perp+(k_T/2)\hat x)^2}{2m} + q\hat F^2_z -\frac{\hbar\delta}{2}
\end{pmatrix}
\begin{pmatrix}
\hat\phi_1({\bf k}_-) \\ \hat\phi_2({\bf k}_+)
\end{pmatrix},
\end{split}
\end{equation}
where $\hat \phi^\dagger_{n}({\bf k}_\perp)=( \phi^\dagger_{n,1} ({\bf k}_\perp), \phi^\dagger_{n,0} ({\bf k}_\perp),  \phi^\dagger_{n,\bar1} ({\bf k}_\perp))$ and ${\bf k}_\pm={\bf k}_\perp \pm (k_T/2) \hat x$. If we choose the detuning to be zero ($\delta=0$), then we can diagonalize this matrix by writing the spin components explicitly.  Scale the eigenenergies by $E_T=\hbar^2 k_T^2/(2m)$, we obtain the final expression
\begin{equation}
 E_{\alpha,\beta}= q+\frac{\hbar^2}{2m}\bigg[ {\bf k}_\perp^2+\frac{1}{4}k^2_T\bigg] \pm \sqrt{\bigg[\frac{\hbar^2}{2m}k_x k_T\bigg]^2 +\Omega^2} ,  E_0 =\frac{\hbar^2 {\bf k}_\perp^2}{2m}
\end{equation}
where $E_\alpha$, $E_\beta$ corresponds to the lower and higher energy band respectively, and $E_0$ is the energy band for $m=0$ spin component.  This is Eq.\ (3) of the main text.

\begin{center}
\vskip 0.3cm
\textbf{Eigenvectors of independent particle Hamiltonian}
\vskip 0.3cm
\end{center}

The eigenvectors of the independent particle Hamiltonian 
${\hat H_{\rm IP}}$, shown in 
Eq.~(4) 
of the main text, are 
\begin{equation}
\label{eqn:eigenstates-supplementary-material}
\begin{pmatrix}
\chi_{a\alpha}({\bf k}_\perp) \\ \chi_{a\beta}({\bf k}_\perp)
\end{pmatrix}
=
\begin{pmatrix}
u_{-\alpha}({\bf k}_\perp) & u_{+\alpha}({\bf k}_\perp)\\ 
u_{-\beta}({\bf k}_\perp) & u_{+\beta}({\bf k}_\perp)
\end{pmatrix}
\begin{pmatrix}
\phi_{1,a}({\bf k}_-) 
\\ 
\phi_{2,a}({\bf k}_+)
\end{pmatrix}
\end{equation}
written as linear combinations of
$\phi_{1,a}({\bf k}_-)$
and 
$\phi_{2,a}({\bf k}_+)$,
where ${\bf k}_{\pm} = {\bf k}_\perp \pm (k_T/2) {\hat {\bf x}}$
are shifted momenta due to the nematic-orbit coupling.
The expressions for the coefficients $u_{\pm\alpha}({\bf k}_\perp)$ snd
$u_{\pm\beta}({\bf k}_\perp)$, that relate the two basis,  are 

\begin{equation}
\label{eqn:u-coefficients-supplementary-material}
\begin{split}
u_{+\alpha}({\bf k}_\perp)
& =
+\frac{1}{\sqrt 2}
\left[ 
1 
- 
f({\widetilde k}_x)
\right]^{1/2}
\quad , \quad
u_{-\alpha}({\bf k}_\perp)
=
-\frac{1}{\sqrt 2}
\left[ 
1 
+ 
f({\widetilde k}_x)
\right]^{1/2}
\\
u_{+\beta}({\bf k}_\perp)
& =
+\frac{1}{\sqrt 2}
\left[ 
1 
+
f({\widetilde k}_x)
\right]^{1/2}
\quad , \quad
u_{-\beta}({\bf k}_\perp)
=
+\frac{1}{\sqrt 2}
\left[ 
1 
- 
f({\widetilde k}_x)
\right]^{1/2},
\end{split}
\end{equation}
where the function 
$
f({\widetilde k}_x) 
= 
{\widetilde k}_x/\sqrt{{\widetilde k}_x^2 + {\widetilde \Omega}^2}
$
is expressed in terms of the dimensionless momentum ${\widetilde k}_x = k_x /k_T$
and nematic-orbit amplitude ${\widetilde \Omega} = \Omega/E_T$ defined
in the main text. Notice that the matrix containing the coefficients
$u_{\pm \alpha}({\bf k}_\perp )$ and $u_{\pm \beta} ({\bf k}_\perp )$ is unitary and
that these coefficients are dimensionless, and depend only on ${\widetilde k}_x$
and ${\widetilde \Omega}$.

\begin{center}
\vskip 0.3cm
\textbf{Bogoliubov spectrum of easy-plane nematic phase in the
single-well regime}
\vskip 0.3cm
\end{center}

To obtain the Bogoliubov spectrum of the easy-plane nematic phase 
in the single-well regime, we start from the  
interaction Hamiltonians ${\hat H}_0$ and ${\hat H}_2$, written 
in momentum space in Eqs.~(5) 
and~(7)
of the main text, and group them together as 
\begin{equation}
\label{eqn:interaction-hamiltonian-supplementary-material}
\begin{split}
\hat H_{int}  
& =
\frac{1}{2L^2_\perp}\sum_{\{n_i\}} C^{n_1,n_2}_{n_3,n_4} \bigg[
\sum_{aa^\prime}\sum_{{\bf k}_\perp, {\bf k^\prime}_\perp, {\bf p}_\perp} 
c_0\phi^\dagger_{n_1,a}({\bf k}_\perp-{\bf p}_\perp/2) 
\phi^\dagger_{n_2,a^\prime}({\bf k^\prime}_\perp+{\bf p}_\perp/2) 
\phi_{n_3,a^\prime}({\bf k^\prime}_\perp-{\bf p}_\perp/2) 
\phi_{n_4,a}({\bf k}_\perp+{\bf p}_\perp/2)\\
& + 
\sum_{a a^\prime b b^\prime}\sum_{{\bf k}_\perp,{\bf k^\prime}_\perp,{\bf p}_\perp} 
c_2\phi^\dagger_{n_1,a} ({\bf k}_\perp-{\bf p}_\perp/2) 
F^\mu_{ab} 
\phi_{n_4,b} ({\bf k}_\perp+{\bf p}_\perp/2) 
\phi^\dagger_{n_2,a^\prime} ({\bf k^\prime}_\perp+{\bf p}_\perp/2) 
F^\mu_{a^\prime b^\prime} 
\phi_{n_3,b^\prime} ({\bf k^\prime}_\perp-{\bf p}_\perp/2) \bigg],
\end{split}
\end{equation}
where $F^\mu_{ab}$ is the matrix element of the spin-1 operator 
with spin components $a b$ in the $\mu$ direction,
$L_\perp$ is the length of real space perpendicular to the trap along $z$, $a, a^\prime, b, b^\prime$ 
represent spin-1 components $\{+1, 0, -1\}$, and $\{n_i\}$ denotes the set of trapped states with 
quantum numbers $(n_1,n_2,n_3,n_4)$. Here, $c_0$ and  $c_2$ are 
the spin-independent and spin-dependent interaction strengths,
respectively. And the coefficients $C^{n_1,n_2}_{n_3,n_4}$ are the modification factors due to integration along $z$ axis and have the following relation,
\begin{equation}
C^{n_1,n_2}_{n_3,n_4}=\int^{L_z/2}_{-L_z/2} \mathrm{d}z \varphi^*_{n_1}(z) \varphi^*_{n_2}(z)\varphi_{n_3}(z)\varphi_{n_4}(z),
\end{equation}
where $\varphi_{n_i}(z)$ is the trap state wave function and  $L_z$ is length of $z$ dimension. To be simply, we choose a box trap as the trap potential. Since $n_i$ is restricted to be $1$ or $2$, the lowest two states, we can write the integral explicitly, 
$C^{n_1,n_2}_{n_1,n_2}=1/L_z (n_1\neq n_2)$ and $C^{n_1,n_1}_{n_1,n_1}=3/(2L_z)$.

For the easy-plane nematic phase, Bose-Einstein condensation 
can occur only in the eigenstates $\chi_{a\alpha}$ or $\chi_{a\beta}$.
Therefore, we rewrite the field operators
$\phi_{n_i,a} ({\bf k}_\perp)$ appearing in 
Eq.~(\ref{eqn:interaction-hamiltonian-supplementary-material})
by inverting the relation displayed
in Eq.~(\ref{eqn:eigenstates-supplementary-material}), leading to
\begin{equation}
\begin{pmatrix}
\phi_{1,a}({\bf k}_-) \\ \phi_{2,a}({\bf k}_+)
\end{pmatrix}
=
\begin{pmatrix}
u_{-\alpha}({\bf k}_\perp) & u_{-\beta}({\bf k}_\perp)\\ 
u_{+\alpha}({\bf k}_\perp) & u_{+\beta}({\bf k}_\perp)
\end{pmatrix}
\begin{pmatrix}
\chi_{a\alpha}({\bf k}_\perp) \\ \chi_{a\beta}({\bf k}_\perp)
\end{pmatrix}
\end{equation}

%
%

In the easy-plane nematic phase, the $\alpha$ band 
has the lowest energy. Bose-Einstein condensation occurs only 
at the minimum of the $\alpha$ band, when the energies of the minima 
in the $a = 0$ and $\beta$ bands are much higher. Therefore,   
condensation involving $\phi_0({\bf k}_\perp)$ and $\chi_{a\beta}({\bf k}_\perp)$
does not occur and the interaction Hamiltonian can be approximated by 
\begin{equation}
\begin{split}
\label{eqn:interaction-hamiltonian-supplementary material}
\hat H_{int}\approx &\sum_{aa'}\frac{c_0+aa'c_2}{2L^2_\perp}\sum_{\{n\}}C^{n_1,n_2}_{n_3,n_4}\sum_{{\bf k}_\perp,{\bf k}'_\perp,{\bf p}_\perp}u^*_{n_1}\bigg({\bf k}_\perp-\frac{{\bf p}_\perp}{2}+(-)^{n_1+1}\frac{k_T}{2}{\hat {\bf x}} \bigg) u^*_{n_2}\bigg({\bf k}'_\perp+\frac{{\bf p}_\perp}{2}+(-)^{n_2+1}\frac{k_T}{2}{\hat {\bf x}} \bigg) \\
&\times u_{n_3}\bigg({\bf k}'_\perp-\frac{{\bf p}_\perp}{2}+(-)^{n_3+1}\frac{k_T}{2}{\hat {\bf x}} \bigg)  u_{n_4}\bigg({\bf k}_\perp+\frac{{\bf p}_\perp}{2}+(-)^{n_4+1}\frac{k_T}{2}{\hat {\bf x}} \bigg)\chi^\dagger_{a\alpha}\bigg({\bf k}_\perp-\frac{{\bf p}_\perp}{2}+(-)^{n_1+1}\frac{k_T}{2}{\hat {\bf x}}\bigg) \\ 
&\times\chi^\dagger_{a'\alpha}\bigg({\bf k}'_\perp+\frac{{\bf p}_\perp}{2}+(-)^{n_2+1}\frac{k_T}{2}{\hat {\bf x}} \bigg) \chi_{a'\alpha}\bigg({\bf k}'_\perp-\frac{{\bf p}_\perp}{2}+(-)^{n_3+1}\frac{k_T}{2}{\hat {\bf x}} \bigg) \chi_{a\alpha}\bigg({\bf k}_\perp+\frac{{\bf p}_\perp}{2}+(-)^{n_4+1}\frac{k_T}{2}{\hat {\bf x}} \bigg)
\end{split}
\end{equation}
provided that one is sufficiently far below the phase boundary line 
${\widetilde q}_c ({\widetilde \Omega})$, indicated
in Fig.~2 of the main text.

Combining the interaction Hamiltonian above with the kinetic energy of
the $\alpha$-band, assuming that condensation occurs only 
in $\chi_{a \alpha} ({\bf k}_\perp)$ at ${\bf k}_\perp = {\bf 0}$, 
and considering that the interaction 
energy is sufficiently small to avoid populating the $a = 0$ 
and $\beta$ bands, we obtain the quadratic Hamiltonian  
\begin{equation}
\label{eqn:single-well-bogoliubov-hamiltonian-supplementary-material}
\begin{split}
\hat H & = G_{\rm sw} + \frac{1}{2}\sum_{\bf k} 
{\bf X}^\dagger_{\bf k}
\begin{pmatrix}
{\bf E}_1 & {\bf D}\\
{\bf D}^\dagger & {\bf E}_{\bar 1}
\end{pmatrix}
{\bf X}_{\bf k},
\end{split}
\end{equation}
describing excitations (fluctuations) above the condensate.
The ground state energy is $G_{\rm sw}$ and
$
{\bf X}_{\bf k}^\dagger
= 
\begin{pmatrix}
\chi^\dagger_{1}({\bf k}_\perp) &\chi_{1}(-{\bf k}_\perp) 
&\chi^\dagger_{\bar1}({\bf k}_\perp) &\chi_{\bar1}(-{\bf k}_\perp)
\end{pmatrix}
$
is a four-dimensional Bogoliubov spinor. Here, we drop the index
$\alpha$ from the notation, because only the $\alpha$ band 
is considered. The block matrices 
for spin-preserving processes are
\begin{equation}
{\bf E}_a
= 
\begin{pmatrix}
E_g({\bf k}_\perp) + c & fe^{i2\Phi_a} \\
fe^{-i2\Phi_a} & E_g({\bf k}_\perp) + c 
\end{pmatrix},
\end{equation}
where $a = \{+1, -1\}$ is represented by $\{1, {\bar 1}\}$,
$\Phi_a$ is the spin-dependent phase of the condensate 
in the $\alpha$-band at ${\bf k}_\perp= {0}$ 
and $c$, $f$ are energy variables proportional to the spin-preserving
interaction energy $(c_0 + c_2)n_c$,
that is, $c = (c_0 + c_2)n_c A_{\alpha} ({\bf k}_\perp)/4$ and $f = (c_0 + c_2)n_c B_{\alpha} ({\bf k}_\perp)/4$, where $n_c$ is the total density. 
The energy 
$
E_g({\bf k}_\perp)
= 
E_\alpha({\bf k}_\perp)- E_\alpha(0)
,
$
where $E_{\alpha} ({\bf k}_\perp)$ is the eigenenergy defined
in Eq.~(3) of the main text,
is a measure of the excitation energy with respect to the minimum
of the $\alpha$-band.
The block matrices for spin-flip processes are 
\begin{equation}
{\bf D}
= 
\begin{pmatrix}
de^{i(\Phi_1-\Phi_{\bar 1})} & ge^{i(\Phi_{\bar 1}+\Phi_1)} \\
ge^{-i(\Phi_1+\Phi_{\bar 1})} & de^{-i(\Phi_1-\Phi_{\bar 1})}\\
\end{pmatrix},
\end{equation}
and ${\bf D}^\dagger$, 
where $d$ and $g$ are energy variables proportional to the spin-flip
interaction energy $(c_0 - c_2)n_c$, 
that is, $d = (c_0 - c_2)n_c A_\alpha ({\bf k}_\perp)/4$ and $g = (c_0 - c_2)n_c B_\alpha ({\bf k}_\perp)/4$. The function
$
A_{\alpha} ({\bf k}_\perp) = 
\left[
5/2 - 2u_{+\alpha} ({\bf k}_\perp) u_{-\alpha} ({\bf k}_\perp)
\right]
$
describes the effects of the nematic-orbit coupling on the interaction
parameters $c$ and $d$, while the function 
$
B_{\alpha} ({\bf k}_\perp) = 
\left[
2 - 3u_{+\alpha} ({\bf k}_\perp) u_{-\alpha} ({\bf k}_\perp)
\right]
$
describes 
the effects of the nematic-orbit coupling on the interaction
parameters $f$ and $g$. Using the expressions for 
$u_{+\alpha} ({\bf k}_\perp)$ and $u_{-\alpha} ({\bf k}_\perp)$ in 
Eq.~(\ref{eqn:u-coefficients-supplementary-material}), we obtain
\begin{equation}
\label{eqn:g-alpha-supplementary-material}
A_{\alpha} ({\bf k}_\perp) = 
\frac{5}{2}+ 
\frac{\vert {\widetilde \Omega} \vert}
{\sqrt { {\widetilde k}_x^2 + {\widetilde \Omega}^2 } },
B_{\alpha} ({\bf k}_\perp) = 
2 + 
\frac{3}{2} 
\frac{\vert {\widetilde \Omega} \vert}
{\sqrt { {\widetilde k}_x^2 + {\widetilde \Omega}^2 } },
\end{equation}
where ${\widetilde \Omega} = \Omega/E_T$ and 
${\widetilde k}_x = k_x/k_T$, as defined in the main text.

The derivation of the Bogoliubov Hamiltonian in 
Eq.~(\ref{eqn:single-well-bogoliubov-hamiltonian-supplementary-material})
takes into account all fluctuation process to quadratic order that
satisfy momentum, spin and energy conservation, but includes only processes
with small momentum transfer, that is, $\vert \Delta {\bf k} \vert < k_T$.
To perform the Bogoliubov transformation and diagonalize the Hamiltonian
in Eq.~(\ref{eqn:single-well-bogoliubov-hamiltonian-supplementary-material}),
while preserving the bosonic commutation relations, it is necessary to 
multiply the $4 \times 4$ matrix containing the block matrices ${\bf E}_1$,
${\bf E}_{\bar 1}$, ${\bf D}$, and ${\bf D}^\dagger$ by the bosonic metric
\begin{equation}
{\bf G}_{\rm sw}
=
\begin{pmatrix}
1 & 0 & 0 & 0 \\
0 & -1 & 0 & 0 \\
0 & 0 & 1 & 0 \\
0 & 0 & 0 & -1
\end{pmatrix}.
\end{equation}
The diagonalization of the resulting matrix can be obtained analytically 
and gives four eigenvalues, two positive and two negative. The negative 
eigenvalues can be turned into positive ones via normal ordering of 
the resulting Bogoliubov operators. The positive eigenvalues are
\begin{equation}
\begin{split}
&\epsilon_{b,1} ({\bf k}_\perp)
=\sqrt{[E_g({\bf k}_\perp) + (c+d)]^2 - (f+g)^2}, \\
&\epsilon_{b,2} ({\bf k}_\perp)
=\sqrt{[E_g({\bf k}_\perp) + (c-d)]^2 - (f-g)^2},
\end{split}
\end{equation}
and describe two linearly dispersing modes at low momentum. The interaction
parameters are 
$(c + d) = c_0 n_c A_{\alpha} ({\bf k}_\perp)/2$, $(f + g) = c_0 n_c B_{\alpha} ({\bf k}_\perp)/2$ and
$(c - d) = c_2 n_c A_{\alpha} ({\bf k}_\perp)/2$, $(f - g) = c_2 n_c B_{\alpha} ({\bf k}_\perp)/2$, 
where $n_c$ is the total density
and $A_{\alpha} ({\bf k}_\perp)$, $B_{\alpha} ({\bf k}_\perp)$ are given in 
Eq.~(\ref{eqn:g-alpha-supplementary-material}).
The energy 
\begin{equation}
E_g ({\bf k}_\perp) 
= 
\frac{\hbar^2 {\bf k}_\perp^2}{2m} 
+ \vert \Omega \vert
-
\sqrt{
\left(
\frac{\hbar^2 k_x k_T}{2m} 
\right)^2
+ 
\Omega^2
}
\end{equation}
can be simplified in the small momentum regime 
${\widetilde k}_x^2 \ll {\widetilde \Omega}^2$ to the
simple quadratic form
\begin{equation} 
E_g ({\bf k}) 
\approx
\frac{\hbar^2 k_x^2}{2m_x}
+ 
\frac{\hbar^2 k_y^2}{2m_y},
\end{equation}
where the effective masses are
$
m_x 
= 
m/
\left[ 
1 - 1/(2{\widetilde \Omega})
\right]
$
and $m_y = m$. 
This shows explicitly that the nematic-orbit coupling produces 
a heavier mass along the $x$-direction in the easy-plane nematic 
single-well phase, giving $m_x > m$ since $\widetilde \Omega > 1/2$ 
in this phase. As a result the linear dispersions of the modes 
at small momenta is anisotropic. 

In the regime of small momenta, $E_g ({\bf k}_\perp) \ll (c+d), (f+g)$ and
$E_g ({\bf k}_\perp) \ll (c-d), (f-g)$, we can simply prove
\begin{equation}
\begin{split}
&E_g({\bf k}_\perp)+c+d-f-g \approx E_g({\bf k}_\perp) +\frac{c_0n_c}{2}\bigg(\frac{\widetilde k_x}{2\widetilde \Omega}\bigg)^2 , \\
&E_g({\bf k}_\perp)+c-d-f+g \approx E_g({\bf k}_\perp) +\frac{c_2n_c}{2}\bigg(\frac{\widetilde k_x}{2\widetilde \Omega}\bigg)^2 
\end{split}
\end{equation}
are quadratic, and the leading term of
\begin{equation}
\begin{split}
&c+d+f+g \approx  \frac{7}{2}c_0n_c+O(k_x^2), \\
&c-d+f-g \approx \frac{7}{2}c_2n_c+O(k_x^2) 
\end{split}
\end{equation}
are constants.
This leads to excitation spectra 
\begin{equation}
\begin{split}
&\epsilon_{b,1} ({\bf k}_\perp)
\approx \sqrt{ (E_g({\bf k}_\perp)+c+d-f-g) (c+d+f+g)}, \\
&\epsilon_{b,2} ({\bf k}_\perp)
\approx \sqrt{( E_g({\bf k}_\perp)+c-d-f+g) (c-d+f-g) }.
\end{split}
\end{equation}
For mode $1$, the excitation
energy along the $x$-direction is
$ 
\epsilon_{b,1} (k_x, 0)
= 
\hbar \vert k_x \vert c_{1 x}
$
with velocity
$$
c_{1 x} = \frac{1}{2} \sqrt{\frac{7 c_0 n_c}{m_x}+\frac{7 c^2_0 n_c^2}{4\hbar^2 k_T^2 \widetilde\Omega^2}}= \frac{1}{2} \sqrt{\frac{7 c_0 n_c}{m_{1x}}},
$$
where $m_{1x}=m_x/\big(1+\frac{m_x c_0n_c}{4\hbar^2 k_T^2 \widetilde \Omega^2}\big)$, while the excitation energy along the $y$-direction is 
$ 
\epsilon_{b,1} (0, k_y)
= 
\hbar \vert k_y \vert c_{1 y}
$
with velocity
$$
c_{1 y} = \frac{1}{2} \sqrt{\frac{7 c_0 n_c}{m}}.
$$
Since $m_{1x} > m$, it is clear that $c_{1 x} < c_{1 y}$, as illustrated in 
Figs.~3a and~3b of the main text.
For mode $2$, the excitation
energy along the $x$-direction is
$ 
\epsilon_{b,2} (k_x, 0)
= 
\hbar \vert k_x \vert c_{2 x}
$
with velocity
$$
c_{2 x} = \frac{1}{2} \sqrt{\frac{7 c_2 n_c}{m_x}+\frac{7 c^2_2 n_c^2}{4\hbar^2 k_T^2 \widetilde\Omega^2}}=\frac{1}{2} \sqrt{\frac{7 c_2 n_c}{m_{2x}}},
$$
where $m_{2x}=m_x/\big(1+\frac{m_x c_2n_c}{4\hbar^2 k_T^2 \widetilde \Omega^2}\big)$, while the excitation energy along the $y$-direction is 
$ 
\epsilon_{b,2} (0, k_y)
= 
\hbar \vert k_y \vert c_{2 y}
$
with velocity 
$$
c_{2 y} = \frac{1}{2} \sqrt{\frac{7 c_2 n_c}{m}}.
$$
Since $m_{2x} > m$, it is clear that $c_{2 x} < c_{2 y}$, as illustrated in 
Figs.~3a and~3b of the main text. 
In deriving the expressions for the linear mode velocities 
we made use of the relation
$\lim_{{\bf k} \to {\bf 0}} A_{\alpha} ({\bf k}) = 7/2$ and $\lim_{{\bf k} \to {\bf 0}} B_{\alpha} ({\bf k}) = 7/2$. Furthermore,
given that $c_0 > c_2 > 0$ for $^{23}{\rm Na}$, the corresponding
velocities for mode 1 are larger than those for mode 2, 
that is, $c_{1x} > c_{2x}$ and $c_{1y} > c_{2y}$, as can be seen 
also in Figs.~3a and~3b of the main text. 

\begin{center}
\vskip 0.3cm
\textbf{Bogoliubov spectrum of easy-plane nematic phase in the 
double-well regime}
\vskip 0.3cm
\end{center}

To obtain the Bogoliubov spectrum of the easy-plane nematic phase 
in the double-well regime, we follow the same steps that lead to the
approximate interaction Hamiltonian described in 
Eq.~(\ref{eqn:interaction-hamiltonian-supplementary material})
above, that is, we consider only the lower energy $\alpha$-band,
that is, we are sufficiently far below the phase boundary line
${\widetilde q}_c ({\widetilde \Omega})$, shown in Fig.~2 
of the main text.
We calculate the Bogoliubov spectrum in the double-well phase, exclusively
in the regime where the interaction energy is sufficiently small that
quasiparticles are excited in the vicinity of the minimum 
of each well, that is, only excitations near momenta 
${\bf k}_\perp= \pm {\bf k}_0$ are considered, where
${\bf k}_0 = (k_0,0)$. In this case, we define 
operators in the left-well $(L)$ and in the right-well $(R)$ as
\begin{equation}
\chi_{a\alpha} ({\bf k}_\perp) =
\chi_{La} ({\bf k}_\perp) \quad (k_x < 0)
\quad {\rm and} \quad
\chi_{a\alpha} ({\bf k}_\perp) =
\chi_{Ra} ({\bf k}_\perp) \quad (k_x > 0),
\end{equation}
where we drop the $\alpha$-band index on the right hand side of
the relation. We can write the operator $\chi_{a\alpha} ({\bf k}_\perp)$ in 
compact notation as 
\begin{equation}
\label{eqn:chi-operator-double-well-supplementary-material}
\chi_{a\alpha} ({\bf k}_\perp)
=
\Theta_{L}(k_x)\chi_{La} ({\bf k}_\perp)
+
\Theta_{R}(k_x)\chi_{Ra} ({\bf k}_\perp),
\end{equation}
where
$
\Theta_{L}(k_x) = \Theta(-k_x)
$
and 
$
\Theta_{R}(k_x) = \Theta(k_x)
$
with $\Theta(k_x)$ being the Heaviside step function. The step
function has the property: $\Theta(k_x) = 0$ when $k_x < 0$, 
$\Theta(0)=\frac{1}{2}$ when $k_x = 0$ and $\Theta(k_x)= 1$ 
when $k_x > 0$. 

We replace the original operators $\chi_{a\alpha}({\bf k}_\perp)$ 
in terms of $\chi_{La} ({\bf k}_\perp)$ and $\chi_{Ra} ({\bf k}_\perp)$ in 
the interaction Hamiltonian of 
Eq.~(\ref{eqn:interaction-hamiltonian-supplementary material}),
add the kinetic energy contribution, assume that Bose-Einstein 
condensation occurs simultaneously in both wells and consider 
only low-momentum-transfer excitation processes  
that conserve momentum, energy and spin. Under these considerations,
the Bogoliubov Hamiltonian becomes
\begin{equation}
\label{eqn:double-well-bogoliubov-hamiltonian-supplementary-material}
\hat H = 
G_{\rm dw} 
+
\frac{1}{2}
\sum_{{\bf k}\neq 0}
{\bf Y}_{\bf k}^\dagger 
\begin{pmatrix}
{\bf M}_{LL} & {\bf M}_{LR} \\
{\bf M}_{RL} & {\bf M}_{RR} 
\end{pmatrix}
{\bf Y}_{\bf k},
\end{equation}
where
$
{\bf Y}^\dagger_{\bf k}
=
\begin{pmatrix}
{\bf X}^\dagger_L ({\bf k}_\perp) &
{\bf X}^\dagger_R ({\bf k}_\perp)
\end{pmatrix}
$
is an eight-dimensional vector with 
four dimensional components 
$
{\bf X}^\dagger_{j} ({\bf k}_\perp)
=
\begin{pmatrix}
\chi^\dagger_{j 1}({\bf k}_\perp \mp {\bf k}_0) & 
\chi_{j 1}(-{\bf k}_\perp \pm {\bf k}_0) & 
\chi^\dagger_{j \bar 1}({\bf k}_\perp \mp {\bf k}_0) & 
\chi_{j \bar 1} (-{\bf k}_\perp \pm {\bf k}_0)
\end{pmatrix}
$
in the $j = \{ L, R \}$ sectors corresponding to the upper and lower sign respectively, 
and $G_{\rm dw}$ is the ground state energy.
The ${\bf M}_{ij}$ matrices describe the intra-well $(i = j)$ and
the inter-well $(i \ne j)$ spin processes and are momentum dependent, that
is ${\bf M}_{ij} = {\bf M}_{ij} ({\bf k}_\perp)$. We do not write explicitly this
momentum dependence to avoid clutter in the notation, but we use 
${\bar {\bf k}} = - {\bf k}$ and $a = \{ 1, {\bar 1} \}$ with ${\bar 1} = -1$ 
to identify momentum and spin dependencies of block matrices within ${\bf M}_{ij}$.

The block matrices describing intra-well processes are   
\begin{equation}
{\bf M}_{LL}
=
\begin{pmatrix}
{\bf E}_{L1} ({\bf k}_\perp)     & {\bf D}_L ({\bf k}_\perp) \\
{\bf D}^\dagger_L ({\bf k}_\perp) & {\bf E}_{L {\bar 1}} ({\bf k}_\perp)
\end{pmatrix}
\quad
{\rm and}
\quad
{\bf M}_{RR}
=
\begin{pmatrix}
{\bf E}_{R1} ({\bar {\bf k}_\perp})     & {\bf D}_R ({\bar {\bf k}_\perp}) \\
{\bf D}^\dagger_R ({\bar {\bf k}_\perp}) & {\bf E}_{R {\bar 1}} ({\bar {\bf k}_\perp})
\end{pmatrix},
\end{equation}
where the block matrices for spin-preserving processes are
\begin{equation}
{\bf E}_{j a} ({\bf k}_\perp) =
\begin{pmatrix}
E^\prime_{{\bf k}} + \eta_{0{\bf k}}+\eta_{2{\bf k}} 
&(\xi_{0{\bf k}}+\xi_{2{\bf k}})e^{i2\Phi_{ja}} \\
(\xi_{0{\bf k}}+\xi_{2{\bf k}})e^{-i2\Phi_{ja}} 
& E^\prime_{\bar{\bf k}}+\eta_{0\bar{\bf k}}+\eta_{2\bar{\bf k}}
\end{pmatrix},
\end{equation}
while the block matrices for spin-flip processes are
\begin{equation}
{\bf D}_j ({\bf k}_\perp)
=
\begin{pmatrix}
(\eta_{0{\bf k}}-\eta_{2{\bf k}})e^{i(\Phi_{j1}-\Phi_{j{\bar 1}})}
& (\xi_{0{\bf k}}-\xi_{2{\bf k}})e^{i(\Phi_{j 1}+\Phi_{j {\bar 1}})} \\
(\xi_{0{\bf k}}-\xi_{2{\bf k}})e^{-i(\Phi_{j 1}+\Phi_{j {\bar 1}})} 
&(\eta_{0\bar{\bf k}}-\eta_{2\bar{\bf k}})
e^{-i(\Phi_{j 1}-\Phi_{ j {\bar 1}})} 
\end{pmatrix} 
\end{equation}
and ${\bf D}_j^\dagger ({\bf k}_\perp)$. 
In order to characterize these matrices fully, 
we identify each entry for every matrix element. 
The factors $\Phi_{ja}$ appearing in matrices
${\bf E}_{ja} ({\bf k}_\perp)$ and ${\bf D}_j ({\bf k}_\perp)$ 
are the phases of the condensates
in well $j = \{ L, R \}$ and spin state $a = \{1, \bar 1 \}$.
The diagonal entries for matrices ${\bf E}_{ja} ({\bf k}_\perp)$ 
are uniquely determined by the function
\begin{equation}
E^\prime_{{\bf k}} = E({\bf k}_\perp)\Theta^2(-k_x + k_0),
\end{equation}
where 
$
E({{\bf k}}_\perp)=E_\alpha({\bf k}_\perp-{\bf k}_0)- E_\alpha(-{\bf k}_0)
$
is expressed in the terms of the $\alpha$-band energies 
\begin{equation}
E_{\alpha} ({\bf k}_\perp)
=
q
+
\frac{\hbar^2}{2m} 
\bigg[ k_\perp^2
+
\frac{k^2_T}{4} \bigg] 
-
\sqrt{
\bigg[
\frac{\hbar^2}{2m}k_ x k_T
\bigg]^2
+
\Omega^2
},
\end{equation}
which contain explicitly the nematic-orbit coupling parameters
$\Omega$ and $k_T$, 
and by the functions
\begin{equation}
\label{eqn:eta-function-supplementary-material}
\eta_{\ell {\bf k}}
=
\frac{c_\ell n_c}{4}
C({\bf k}_\perp) 
\Theta^2(- k_x +  k_0),
\end{equation}
where $\ell = \{0, 2\}$ labels the interaction contribution from 
$c_0$ and $c_2$, $n_c$ is the particle density, and 
\begin{equation}
\begin{split}
C({\bf k}_\perp)
&=
\frac{3}{2}\bigg[ u^2_{+\alpha}(-{\bf k}_0)u^2_{+\alpha}({\bf k}_\perp-{\bf k}_0)  +u^2_{-\alpha}(-{\bf k}_0)u^2_{-\alpha}({\bf k}_\perp-{\bf k}_0)\bigg]
+
\bigg[  u^2_{+\alpha}({\bf k}_0)u^2_{+\alpha}({\bf k}_\perp-{\bf k}_0)  +u^2_{-\alpha}({\bf k}_0)u^2_{-\alpha}({\bf k}_\perp-{\bf k}_0) \\
&+2u_{+\alpha}({\bf k}_0)
u_{-\alpha}({\bf k}_0)
u_{-\alpha}({\bf k}_\perp-{\bf k}_0)
u_{+\alpha}({\bf k}_\perp-{\bf k}_0)
\bigg]
\end{split}
\end{equation}
is a coherence factor containing the amplitudes defined
in Eq.~(\ref{eqn:eigenstates-supplementary-material}).
The off-diagonal entries for matrices ${\bf E}_{ja}({\bf k}_\perp)$ 
are uniquely determined by the function
\begin{equation}
\label{eqn:xi-function-supplementary-material}
\xi_{\ell {\bf k}}
=
\frac{c_\ell n_c}{4}A({\bf k}_\perp)
\Theta(k_x + k_0)\Theta(-k_x + k_0),
\end{equation}
where $c_{\ell}$ is either $c_0$ or $c_2$, $n$ is the particle density
and 
\begin{equation}
\begin{split}
A({\bf k}_\perp)
&
=\frac{3}{2} \bigg[u_{-\alpha}(-{\bf k}_0) u_{-\alpha}(-{\bf k}_0)u_{-\alpha}(-{\bf k}_\perp-{\bf k}_0) u_{-\alpha}({\bf k}_\perp-{\bf k}_0) + u_{+\alpha}(-{\bf k}_0) u_{+\alpha}(-{\bf k}_0)u_{+\alpha}(-{\bf k}_\perp-{\bf k}_0) u_{+\alpha}({\bf k}_\perp-{\bf k}_0)
\bigg] \\
&+2\bigg[u_{+\alpha}(-{\bf k}_0) u_{-\alpha}(-{\bf k}_0)u_{-\alpha}(-{\bf k}_\perp-{\bf k}_0) u_{+\alpha}({\bf k}_\perp-{\bf k}_0) + u_{-\alpha}(-{\bf k}_0) u_{+\alpha}(-{\bf k}_0)u_{+\alpha}(-{\bf k}_\perp-{\bf k}_0) u_{-\alpha}({\bf k}_\perp-{\bf k}_0)
\bigg]
\end{split}
\end{equation}
is a coherence factor containing the amplitudes defined
in Eq.~(\ref{eqn:eigenstates-supplementary-material}).
All the entries for block matrix ${\bf D}_j ({\bf k}_\perp)$ are defined
in terms of the phase factors $\Phi_{j a}$ of the condensates and the functions
$\gamma_{\ell {\bf k}}$ and $\xi_{\ell {\bf k}}$ defined in 
Eqs.~(\ref{eqn:gamma-function-supplementary-material}) 
and~(\ref{eqn:xi-function-supplementary-material}), respectively.

The block matrices describing inter-well processes are
\begin{equation}
{\bf M}_{LR}
=\begin{pmatrix}
{\bf F}_{1} ({\bf k}_\perp) & {\bf C}_{1\bar1} ({\bf k}_\perp)\\
{\bf C}_{{\bar 1} 1} ({\bf k}_\perp) &{\bf F}_{\bar 1} ({\bf k}_\perp)
\end{pmatrix}
\quad
{\rm and}
\quad
{\bf M}_{RL}
= 
{\bf M}_{LR}^\dagger,
\end{equation}
where the block matrices for spin-preserving processes are
\begin{equation}
{\bf F}_{a} ({\bf k}_\perp) =
\begin{pmatrix}
(\zeta_{0{\bf k}}+\zeta_{2{\bf k}})e^{i(\Phi_{La}-\Phi_{Ra})} 
&(\gamma_{0{\bf k}}+\gamma_{2{\bf k}})e^{i(\Phi_{La}+\Phi_{Ra})}  \\
(\gamma_{0\bar{\bf k}}+\gamma_{2\bar{\bf k}})e^{-i(\Phi_{La}+\Phi_{Ra})} 
&(\zeta_{0{\bf k}}+\zeta_{2{\bf k}})e^{i(\Phi_{Ra}-\Phi_{La})}
\end{pmatrix},
\end{equation}
while the block matrices for spin-flip processes are
\begin{equation}
{\bf C}_{a {\bar a}} ({\bf k}_\perp) =
\begin{pmatrix}
(\zeta_{0{\bf k}}-\zeta_{2{\bf k}})e^{i(\Phi_{La}-\Phi_{R\bar a})} 
& (\gamma_{0{\bf k}}-\gamma_{2{\bf k}})e^{i(\Phi_{La}+\Phi_{R\bar a})} \\
(\gamma_{0\bar{\bf k}}-\gamma_{2\bar{\bf k}})e^{-i(\Phi_{La}+\Phi_{R\bar a})} 
& (\zeta_{0{\bf k}}-\zeta_{2{\bf k}})e^{i(\Phi_{R\bar a}-\Phi_{La})} 
\end{pmatrix}.
\end{equation}
The diagonal matrix elements of ${\bf F}_{a} ({\bf k})$
and ${\bf C}_{a {\bar a}} ({\bf k})$ are specified 
in terms of the phase factors $\Phi_{j a}$ of the condensates and the function
$\gamma_{\ell {\bf k}}$,
\begin{equation}
\label{eqn:gamma-function-supplementary-material}
\gamma_{\ell {\bf k}}
=
\frac{c_\ell n_c}{4}
B({\bf k}_\perp) 
\Theta^2(- k_x +  k_0),
\end{equation}
where $\ell = \{0, 2\}$ labels the interaction contribution from 
$c_0$ and $c_2$, $n_c$ is the particle density, and 
\begin{equation}
B({\bf k}_\perp)=1+3u_{+\alpha}({\bf k}_0)u_{-\alpha}({\bf k}_0)u_{-\alpha}({\bf k}_\perp-{\bf k}_0)u_{+\alpha}({\bf k}_\perp-{\bf k}_0)
\end{equation}
is a coherence factor containing the amplitudes defined
in Eq.~(\ref{eqn:eigenstates-supplementary-material}).
The off-diagonal entries for  ${\bf F}_{a} ({\bf k}_\perp)$
and ${\bf C}_{a {\bar a}}({\bf k}_\perp)$ are determined by the function 
\begin{equation}
\zeta_{\ell {\bf k}} = \frac{c_\ell n_c}{4}D({\bf k}_\perp)
\Theta(k_x + k_0)\Theta(-k_x + k_0),
\end{equation} 
where $c_{\ell}$ is either $c_0$ or $c_2$, $n_c$ is the particle density
and 
\begin{equation}
\begin{split}
D({\bf k}_\perp)
&
=\frac{3}{2} \bigg[u_{-\alpha}(-{\bf k}_0) u_{-\alpha}({\bf k}_0)u_{-\alpha}(-{\bf k}_\perp+{\bf k}_0) u_{-\alpha}(-{\bf k}_\perp-{\bf k}_0) + u_{+\alpha}(-{\bf k}_0) u_{+\alpha}({\bf k}_0)u_{+\alpha}(-{\bf k}_\perp+{\bf k}_0) u_{+\alpha}(-{\bf k}_\perp-{\bf k}_0)
\bigg] \\
&+\bigg[u_{+\alpha}(-{\bf k}_0) u_{+\alpha}({\bf k}_0)u_{-\alpha}(-{\bf k}_\perp+{\bf k}_0) u_{-\alpha}(-{\bf k}_\perp-{\bf k}_0) + u_{-\alpha}(-{\bf k}_0) u_{+\alpha}({\bf k}_0)u_{+\alpha}(-{\bf k}_\perp+{\bf k}_0) u_{-\alpha}(-{\bf k}_\perp-{\bf k}_0) \\
&+u_{+\alpha}(-{\bf k}_0) u_{-\alpha}({\bf k}_0)u_{+\alpha}(-{\bf k}_\perp-{\bf k}_0) u_{-\alpha}(-{\bf k}_\perp+{\bf k}_0) + u_{-\alpha}(-{\bf k}_0) u_{-\alpha}({\bf k}_0)u_{+\alpha}(-{\bf k}_\perp+{\bf k}_0) u_{+\alpha}(-{\bf k}_\perp-{\bf k}_0) 
\bigg]
\end{split}
\end{equation}
is a coherence factor containing the amplitudes defined
in Eq.~(\ref{eqn:eigenstates-supplementary-material}).

The eigenvalues of the $8 \times 8$ Bogoliubov matrix 
containing the block matrices ${\bf M}_{ij}$ 
in Eq.~(\ref{eqn:double-well-bogoliubov-hamiltonian-supplementary-material}),
are obtained by performing a Bogoliubov transformation that diagonalizes 
the Hamiltonian while preserving the bosonic commutation relations. 
For this purpose, we use the metric matrix
\begin{equation}
{\bf G}_{\rm dw} 
=
\begin{pmatrix}
1 & 0 & 0 & 0 & 0 & 0 & 0 & 0 \\
0 & -1 & 0 & 0 & 0 & 0 & 0 & 0 \\
0 & 0 & 1 & 0 & 0 & 0 & 0 & 0 \\
0 & 0 & 0 & -1 & 0 & 0 & 0 & 0 \\
0 & 0 & 0 & 0 & 1 & 0 & 0 & 0 \\
0 & 0 & 0 & 0 & 0 & -1 & 0 & 0 \\
0 & 0 & 0 & 0 & 0 & 0 & 1 & 0 \\
0 & 0 & 0 & 0 & 0 & 0 & 0 & -1 \\
\end{pmatrix},
\end{equation}
and obtain the eight eigenvalues numerically. As expected four eigenvalues are
positive and four are negative, but the negative eigenvalues 
can be made positive by a particle-hole transformation. 
Thus, in Fig.~3 of the main text, we plot the dispersion of the
four collective modes found and indicate that all four of them are linear at low momenta. All the modes are affected by the 
nematic-orbit coupling as discussed in the main text, where we also provide a 
qualitative analysis of the nature of the modes based on in-phase and out-phase 
relations of the corresponding eigenvectors of the Bogoliubov matrix.
The building block of the analysis of the modes is that, if there were no 
spin-spin interactions, the double-well for spins $1$ and ${\bar 1}$ would be 
independent from each other. This means that each independent system would exhibit two
linear modes, which would be the same for spin components
$1$ and ${\bar 1}$. However, when spin-spin interactions are included, the degeneracy
of the modes is lifted, resulting into four split linear modes.

\begin{center}
\vskip 0.3cm
\textbf{Real space description of easy-plane nematic phases: effective Hamiltonian}
\vskip 0.3cm
\end{center}

For easy-plane nematic phases with zero magnetization, 
the density of particles in the $a = 0$ spin state is $n_0 = 0$,
while for spin states $a = \{ 1, {\bar 1} \}$ is $n_{1} = n_{\bar 1} \ne 0$.  
Sufficiently far below the phase boundary 
${\widetilde q}_{c} ({\widetilde \Omega})$ shown in Fig.~2 
of the main text, the only available spin states are $a = \{1, {\bar 1} \}$.
In this regime, the total Hamiltonian simplifies to
\begin{equation}
\begin{split}
\label{eqn:easy-plane-hamiltonian-supplementary-material}
\hat H=
\int \mathrm{d}^2 r_\perp
\begin{pmatrix}
\hat \psi_1^\dagger({\bf r}_\perp) & \hat \psi_2^\dagger({\bf r}_\perp)
\end{pmatrix}
\begin{pmatrix}
\frac {{\bf p}_\perp^2}{2m} + q\hat F^2_z  & \Omega e^{-i k_T x}\hat F^2_z \\
\Omega e^{i k_T x}\hat F^2_z & \frac {{\bf p}_\perp^2}{2m} + q\hat F^2_z 
\end{pmatrix}
\begin{pmatrix}
\hat \psi_1({\bf r}_\perp) \\ \hat \psi_2({\bf r}_\perp)
\end{pmatrix}+\hat H_{int},
\end{split}
\end{equation}
where the real space representation of the interaction part is 
\begin{equation}
\hat H_{int}
=\int  \mathrm{d}^3  r 
\bigg [ 
\sum_{a a^\prime}\frac{c_0}{2}
\psi^\dagger_a({\bf r})  \psi^\dagger_{a^\prime}({\bf r})  
\psi_{a^\prime}({\bf r})  \psi_a({\bf r})  
+ 
\sum_{a a^\prime b b^\prime}\frac{c_2}{2}\psi^\dagger_a({\bf r})  
\psi^\dagger_{a^\prime}({\bf r})  
{\hat{\bf F}}_{ab} \cdot {\hat{\bf F}}_{a^\prime b^\prime}
\psi_{b^\prime}({\bf r})  \psi_b({\bf r}) \bigg ],
\end{equation}
with the summation over spin indices including only states $\{ 1, {\bar 1} \}$ 
and it can be proved to be invariant under rotating wave approximation(RWA).
The interaction term can then be simplified to 
\begin{equation}
\begin{split}
\hat H_{int} =
\int  \mathrm{d}^3  r
\bigg [\sum_{aa^\prime}
\frac{c_0}{2}
\psi^\dagger_a ({\bf r}) \psi^\dagger_{a^\prime}({\bf r})  
\psi_{a^\prime}({\bf r}) \psi_a({\bf r}) 
+
\sum_{aa^\prime}
\frac{c_2}{2} aa^\prime 
\psi^\dagger_a({\bf r})    \psi^\dagger_{a^\prime}({\bf r})  
\psi_{a^\dagger}({\bf r})  \psi_a({\bf r})  \bigg ].
\end{split}
\end{equation}

In the mean-field approximation, we replace the operators 
$\psi_a^\dagger ({\bf r})$ and $\psi_a ({\bf r})$ by the 
condensate wave functions $\psi_a^* ({\bf r})$ and $\psi_a ({\bf r})$, 
$\psi_{n,a}^\dagger ({\bf r}_\perp)$ and $\psi_{n,a} ({\bf r}_\perp)$ by 
the wave functions $\psi_{n,a}^* ({\bf r}_\perp)$ and $\psi_{n,a} ({\bf r}_\perp)$  
and write the effective Hamiltonian for the easy-plane nematic phase as
\begin{equation}
\hat H_{\rm EP}
= 
\int \mathrm{d}^2 r_\perp
\begin{pmatrix}
\hat \psi_1^*({\bf r}_\perp) & \hat \psi_2^*({\bf r}_\perp)
\end{pmatrix}
\begin{pmatrix}
\frac {{\bf p}_\perp^2}{2m} + q\hat F^2_z  & \Omega e^{-i k_T x}\hat F^2_z \\
\Omega e^{i k_T x}\hat F^2_z & \frac {{\bf p}_\perp^2}{2m} + q\hat F^2_z 
\end{pmatrix}
\begin{pmatrix}
\hat \psi_1({\bf r}_\perp) \\ \hat \psi_2({\bf r}_\perp)
\end{pmatrix}
+
{\hat H}_{\rm I}
\end{equation}
with ${\hat \psi}^*_{n}({\bf r}_\perp)=(\psi^*_{n,1}({\bf r}_\perp),\psi^*_{n,0}({\bf r}_\perp),\psi^*_{n,{\bar 1}}({\bf r}_\perp))$.
The interaction Hamiltonian is now
\begin{equation}
{\hat H}_{\rm I}
=
\int  \mathrm{d}^3 r 
\bigg[
\frac{c_0}{2}
\bigg( |\psi_1({\bf r})|^2  + |\psi_{\bar 1}({\bf r})|^2 \bigg)^2 
+ 
\frac{c_2}{2}
\bigg( |\psi_1({\bf r})|^2 - |\psi_{\bar 1}({\bf r})|^2 \bigg)^2 
\bigg] ,
\end{equation}
with $c_0 > c_2 >0$ as in $^{23}{\rm Na}$. 
Since the spin-spin interactions are antiferromagnetic $(c_2 > 0)$, 
the interaction energy is minimized when the local condensate densities 
are the same, that is, 
$
\vert \psi_{1}({\bf r}) \vert^2 
= 
\vert \psi_{\bar 1}({\bf r}) \vert^2.
$

\begin{center}
\vskip 0.3cm
\textbf{Real space description of easy-plane nematic phases: single-well regime}
\vskip 0.3cm
\end{center}

In the single-well regime, Bose-condensation occurs at ${\bf k} = 0$ 
in the $\alpha$-band, that is, the $\chi_{a \alpha} ({\bf k})$ operators 
become delta functions $C_a \delta ({\bf k})$ in mean field.  
Neglecting the $\beta$-band in 
Eq.~(\ref{eqn:eigenstates-supplementary-material}) 
and using the mean-field relations
\begin{equation}
\chi_{a \alpha} ({\bf k}_\perp - \frac{k_T}{2}{\hat {\bf x}})
\to 
C_{a}
\delta(k_x-\frac{k_T}{2})\delta(k_y)
\quad 
{\rm and}
\quad
\chi_{a \alpha} ({\bf k}_\perp + \frac{k_T}{2}{\hat {\bf x}})
\to 
C_{a} e^{-i\vartheta}
\delta(k_x + \frac{k_T}{2})\delta(k_y),
\end{equation}
where $\vartheta$ is 
the phase difference between the dressed state condensates,
leads to the momentum space condensate wavefunction 
\begin{equation}
{\bf \Phi}_a({\bf k}_\perp,z)=C_a
\sum_{j=1,2} \phi_{j,a}({\bf k}_\perp)\varphi_j(z)
=C_a
\bigg[u_{-\alpha}({\bf k}_+)\delta(k_x+\frac{k_T}{2})\delta(k_y)\varphi_1(z) + e^{-i\vartheta}u_{+\alpha}({\bf k}_-)\delta(k_x-\frac{k_T}{2})\delta(k_y)\varphi_2(z)\bigg].
\end{equation}
Performing the Fourier transformation
$
\psi_{n,a} ({\bf r}_\perp)
=
\frac{1}{L_\perp}
\sum_{{\bf k}_\perp} 
\phi_{n, a}({\bf k}_\perp)e^{i{\bf k}_\perp \cdot{\bf r}_\perp}
$
in the continuum limit, where 
$
\sum_{{\bf k}_\perp}
\to \left[ 
L^2_\perp/(2\pi)^2
\right]
\int \mathrm{d}^2 k_\perp,
$
and using the relations $u_{+ \alpha} (0) = 1/\sqrt{2}$
and $u_{- \alpha} (0) = - 1/\sqrt{2}$,
leads to the real space wavefunction
\begin{equation}
\begin{split}
\label{eqn:easy-plane-single-well-wavefunction-supplementary-material}
{\bf \Psi}_a({\bf r})=C_a
\sum_{j=1,2} \psi_{j,a}({\bf r}_\perp)\varphi_j(z)
=
{\cal A}_{\rm sw}
e^{-i\frac{\vartheta}{2}}\bigg[
-e^{i\frac{\vartheta}{2}}e^{-i\frac{k_T}{2}x}\varphi_1(z) + e^{-i\frac{\vartheta}{2}}e^{i\frac{k_T}{2}x}\varphi_2(z) \bigg].
\end{split}
\end{equation}
The wavefunction above is the relation displayed in 
Eq.~(16) of the main text,
where the constant 
\begin{equation}
{\cal A}_{\rm sw} = 
\frac{L_\perp}{(2\pi)^2}
\frac{C_a}{\sqrt 2}
\end{equation}
is independent of the spin index $a$, because the condensates for
$a = 1$ and $a = {\bar 1}$ have the same strength $C_a$ in 
the easy-plane nematic phase with zero magnetization, that is,
$C_{1} = C_{\bar 1} = C_{\rm sw}$. Notice that $C_a = C_{\rm sw}$ has
dimensions of $L_\perp^{-2}$, while ${\cal A}_{\rm sw}$ has dimensions of $L_\perp^{-1}$ 
and represents the amplitude of the condensate wavefunction.

The total condensate density for the easy-plane nematic phase in the single-well
regime is (assume $\varphi_j(z)$ is real function)
\begin{equation}
n_C ({\bf r})
=\sum_{a=\pm1}|{\bf \Psi}_{a}({\bf r})|^2
=\sum_{\substack{a=\pm1 \\ j=1,2}}
|\psi_{j,a}({\bf r}_\perp)\varphi_j(z)|^2
=
2|{\cal A}_{\rm sw}|^2[|\varphi_1(z)|^2
+
|\varphi_2(z)|^2
-
2\varphi_1(z)\varphi_2(z)\cos (k_T x -\vartheta)],
\end{equation}
when expressed in terms of trapped states wavefunction $\varphi_n(z)$ and amplitude ${\cal A}_{\rm sw}$. The amplitude ${\cal A}_{\rm sw}$ is found 
by normalizing the condensate density $n_C ({\bf r})$ to the total
number of particles in the condensate
\begin{equation}
N_C
=
\sum_a\int \mathrm{d}^3 r |{\bf \Psi}_{a}({\bf r})|^2,
\end{equation}
and thus the wavefunction amplitude is 
$
\vert {\cal A}_{\rm sw} \vert = 
\sqrt{\frac{N_C}{4L_\perp^2}}.
$
The use of this result for ${\cal A}_{\rm sw}$ in combination with 
the trapped states $\varphi_n(z)$ in the infinitely deep box potential, 
leads to the condensate density
\begin{equation}
\begin{split}
n_C ({\bf r}) 
&=\frac{N_C}{L^2_\perp L_z} \bigg[ \cos^2 \bigg(\frac{\pi}{L_z}z\bigg) + \sin^2 \bigg(\frac{2\pi}{L_z}z\bigg) - 2\cos \bigg(\frac{\pi}{L_z}z\bigg) \sin \bigg(\frac{2\pi}{L_z}z\bigg)\cos (k_T x -\vartheta)  \bigg] \\
&=
\frac{N_C}{V} 
\bigg[ \cos^2 \bigg(\frac{\pi}{L_z}z\bigg) 
+ 
\sin^2 \bigg(\frac{2\pi}{L_z}z\bigg) - 2\cos \bigg(\frac{\pi}{L_z}z\bigg) \sin \bigg(\frac{2\pi}{L_z}z\bigg)\cos (k_T x -\vartheta)\bigg],
\end{split}
\end{equation}
where $V=L^2_\perp L_z$ is defined as the volume of the condensate.
Finally, defining the scaled local condensate density as
$
{\widetilde n}_C ({\bf r}) = n_C ({\bf r})/n_c,
$
where $n_c$ is the total density, results in
\begin{equation}
\label{eqn:easy-plane-single-well-condensate-density-supplementary-material}
{\widetilde  n}_C ({\bf r}) 
= 
\sigma 
\bigg[ \cos^2 \bigg(\frac{\pi}{L_z}z\bigg) 
+ 
\sin^2 \bigg(\frac{2\pi}{L_z}z\bigg) - 2\cos \bigg(\frac{\pi}{L_z}z\bigg) \sin \bigg(\frac{2\pi}{L_z}z\bigg)\cos (k_T x -\vartheta)\bigg],
\end{equation}
where $\sigma = N_C/N$ is the condensate fraction, with $N_C$ being the number
of particles in the condensate and $N$ being the total number of particles. 
Substitute the wavefunction in Eq.~(\ref{eqn:easy-plane-single-well-wavefunction-supplementary-material}) into the full effective Hamiltonian 
and minimize the energy with respect to $\vartheta$, we can fix the random relative phase $\vartheta$ in the density profile. Then the independent particle energy is
\begin{equation}
E_{IP}(\vartheta)=\frac{N_C}{2}\bigg(\frac{\hbar^2 k^2_T}{8m} + q-\Omega\cos \vartheta\bigg),
\end{equation}
and the interaction energy becomes
\begin{equation}
\begin{split}
E_I(\vartheta)&= \int \mathrm{d}^3 r \frac{c_0}{2}n^2_C({\bf r})=\frac{c_0N^2_C}{2V^2} \int \mathrm{d}^3 r  
\bigg[ \cos^2 \bigg(\frac{\pi}{L_z}z\bigg) 
+ 
\sin^2 \bigg(\frac{2\pi}{L_z}z\bigg) - 2\cos \bigg(\frac{\pi}{L_z}z\bigg) \sin \bigg(\frac{2\pi}{L_z}z\bigg)\cos (k_T x -\vartheta)\bigg]^2 \\
&=\frac{c_0 N^2_C}{2V}\bigg( \frac{7}{4} +\frac{\sin k_TL_\perp}{2 k_T L_\perp}\cos 2\vartheta \bigg)
\end{split}
\end{equation}
and total energy becomes 
\begin{equation}
E(\vartheta)=\frac{c_0 N^2_C}{2V}\bigg( \frac{7}{4} +\frac{\sin k_TL_\perp}{2 k_T L_\perp}\cos 2\vartheta \bigg) +\frac{N_C}{2}\bigg(\frac{\hbar^2 k^2_T}{8m} + q-\Omega\cos \vartheta\bigg)
\end{equation}
in combination of the independent particle energy term.
Minimize $E(\vartheta)$ by taking the derivative with respect to $\vartheta$ and end up with 
\begin{equation}
\bigg( \frac{N_C\Omega}{2}-\frac{c_0N^2_C\sin k_TL_\perp}{Vk_TL_\perp}\cos\vartheta\bigg) \sin \vartheta = 0,
\end{equation}
indicating $\vartheta= 0,\pi$, since $\vert N_C\Omega/2 \vert \gg \vert (c_0N^2_C/V k_T L_\perp)\sin k_TL_\perp \cos \vartheta \vert$ since $\vartheta$ is real. Then we take the second derivative of $E(\vartheta)$ and get $\partial^2_\vartheta E(\vartheta)\vert_{\vartheta=0} >0$, therefore the relative phase between two trapped states $\vartheta$ should be equal to $0$ in order to minimize the total energy.

Since the density profile along $z$ direction is only originated from the box trap, 
we can integrate out $z$ direction and scale the density to a 2-dimensional condensate density $n_\perp= N/L^2_\perp$,
\begin{equation}
{\widetilde n}_C({\bf r}_\perp) = \sigma, 
\end{equation}
where ${\widetilde n}_C({\bf r}_\perp)=\int {\widetilde n}_C({\bf r})  \mathrm{d}z /n_\perp$.
This density profile describes a constant density along nematic-orbit coupling axis.
For fixed values of the interaction parameters $c_0$ and $c_2$, 
$\sigma$ is a function of ${\widetilde q}$ and 
${\widetilde \Omega}$. The condensate fraction $\sigma$ for 
the easy-plane nematic single-well phase tends to zero when the phase
boundary ${\widetilde q}_{c} ({\widetilde \Omega})$ is approached
(see Fig.~2 of the main text), since for 
$ {\widetilde q} > {\widetilde q}_{c} ({\widetilde \Omega})$ the
easy-axis nematic phase takes over. However, we use the value of 
$\sigma = 0.7$ for ${\widetilde \Omega} = 1$ in the single-well case, 
because we discussed here only an approximate real space Hamiltonian 
for easy-plane nematic phases far below the phase boundary 
${\widetilde q}_{c} ({\widetilde \Omega})$, where the condensate 
fraction is closer to one. 
 
%
%

\begin{center}
\vskip 0.3cm
\textbf{Real space description of easy-plane nematic phases: double-well regime}
\vskip 0.3cm
\end{center}

In the double-well regime, Bose-condensation occurs simultaneously at 
the right $(R)$ and left $(L)$ wells, that is, at momenta 
${\bf k}_\perp = \pm {\bf k}_0$ of the $\alpha$-band, 
with ${\bf k}_0 = k_0 {\hat {\bf x}}$. This implies that the expectation value of the $\chi_{a \alpha} ({\bf k}_\perp)$ operator 
become a sum of weighted delta functions 
$
C_{a R} \delta ({\bf k}_\perp - {\bf k}_0) 
+ C_{aL} e^{-i \vartheta_{LR}} \delta ({\bf k}_\perp + {\bf k}_0 )
$ 
in mean field. Here, $C_{a R}$ and $C_{a L}$ are the amplitudes 
of the condensates in the right and left wells, and $\vartheta_{LR}$ 
is the phase difference between the condensates in the right and left wells.  
Neglecting the $\beta$-band in 
Eq.~(\ref{eqn:eigenstates-supplementary-material}) 
and using the mean-field relations 
\begin{equation}
\begin{split}
\chi_{a \alpha} ({\bf k} - \frac{k_T}{2}{\hat {\bf x}})
\to
&
\left[
C_{aR}
\delta(k_x - k_0 -\frac{k_T}{2})\delta(k_y)
+ 
C_{aL}
e^{- i \vartheta_{LR}}
\delta(k_x + k_0 -\frac{k_T}{2})\delta(k_y)
\right]
\\
\chi_{a \alpha} ({\bf k} + \frac{k_T}{2}{\hat {\bf x}})
\to 
&
\left[
C_{aR} 
\delta(k_x - k_0 + \frac{k_T}{2})\delta(k_y)
+
C_{aL}
e^{-i\vartheta_{LR}}
\delta(k_x + k_0 + \frac{k_T}{2})\delta(k_y)
\right]
e^{-i\vartheta},
\end{split}
\end{equation}
where $\vartheta$ is the phase difference between the dressed state condensates, 
leads to the momentum space condensate wavefunction 
\begin{equation}
\begin{split}
{\bf \Phi}_a({\bf k}_\perp,z)=
\sum_{j=1,2}
\phi_{j,a}({\bf k}_\perp)\varphi_j(z) &=
u_{-\alpha}({\bf k}_+)\bigg[C_{aR}\delta(k_x-k_0+\frac{k_T}{2})+e^{-i\vartheta_{LR}}C_{aL}\delta(k_x+k_0+\frac{k_T}{2})\bigg]\delta(k_y)\varphi_1(z) \\
&+e^{-i\vartheta}u_{+\alpha}({\bf k}_-)\bigg[C_{aR}\delta(k_x-k_0-\frac{k_T}{2})+e^{-i\vartheta_{LR}}C_{aL}\delta(k_x+k_0-\frac{k_T}{2})\bigg]\delta(k_y)\varphi_2(z)
\end{split}
\end{equation}
Since the left and right wells are perfectly symmetric, the amplitudes
$C_{aL}$ and $C_{aR}$ are identical, that is, $C_{aL} = C_{aR} = C_a$.

Performing the Fourier transformation
$
\psi_{n,a} ({\bf r}_\perp)
=
\frac{1}{L_\perp}
\sum_{{\bf k}_\perp} 
\phi_{n, a}({\bf k}_\perp)e^{i{\bf k}_\perp \cdot{\bf r}_\perp}
$
in the continuum limit, where the summation over momentum states
${\bf k}_\perp$ becomes the integral
$
\left[
L^2_\perp/(2\pi)^2
\right]
\int \mathrm{d}^2  k_\perp,
$
leads to the wave function
\begin{equation}
\begin{split}
{\bf \Psi}_a({\bf r})=C_a
\sum_{j=1,2}\psi_{j,a}({\bf r}_\perp)\varphi_j(z) 
&=\frac{L_\perp}{(2\pi)^2}C_a \bigg[
(u_{-\alpha}({\bf k}_0)e^{i(k_0-\frac{k_T}{2})x} + e^{-i\vartheta_{LR}}u_{-\alpha}(-{\bf k}_0)e^{-i(k_0+\frac{k_T}{2})x})\varphi_1(z) \\
&+ e^{-i\vartheta}(u_{+\alpha}({\bf k}_0)e^{i(k_0+\frac{k_T}{2})x}+e^{-i\vartheta_{LR}}u_{+\alpha}(-{\bf k}_0)e^{-i(k_0-\frac{k_T}{2})x})\varphi_2(z)
\bigg] \\
&=\frac{L_\perp}{(2\pi)^2}C_a e^{-i\frac{\vartheta+\vartheta_{LR}}{2}} 
\sum_{\substack{j=\pm \\ l=\pm}}
\bigg[u_{j\alpha}(l k_0) e^{i\big[(lk_0+j\frac{k_T}{2})x-(j\frac{\vartheta}{2}-l\frac{\vartheta_{LR}}{2}) \big]}\bigg]\varphi_j(z)
\label{eqn:easy-plane-double-well-wavefunction-supplementary-material}
\end{split}
\end{equation}
displayed in Eq.~(17) of the main text.
Here, we denote $\varphi_-(z)=\varphi_1(z)$ and $\varphi_+(z)=\varphi_2(z)$ for simplicity.
Notice that coefficient in front of the brackets of the previous expression 
is independent of the spin state $a$ for easy-plane nematic phase 
since $C_{1} = C_{\bar 1} = C_{\rm dw}$, and thus can be written as  
\begin{equation}
{\cal B}_{\rm dw} 
=
\frac{L_\perp}{(2\pi)^2}C_a.
\end{equation}

The constant ${\cal B}_{\rm dw}$ can be determined by requiring
that the condensate density 
$
n_C ({\bf r}) 
= 
\vert{\bf \Psi}_1 ({\bf r}) \vert^2
+ 
\vert{\bf \Psi}_{\bar 1} ({\bf r}) \vert^2
$
is normalized to $N_C$, which is the total number of condensed particles in 
easy-plane nematic double-well phase. Given that the condensate density is
\begin{equation}
\begin{split}
n_C ({\bf r})
&=
2 \vert {\cal B}_{\rm dw}\vert^2
\Bigg\{\bigg[u^2_{-\alpha}({\bf k}_0)+u^2_{-\alpha}(-{\bf k}_0)+2u_{-\alpha}({\bf k}_0)u_{-\alpha}(-{\bf k}_0)\cos(2 k_0 x+\vartheta_{LR})\bigg]|\varphi_1(z)|^2 \\
&+\bigg[u^2_{+\alpha}({\bf k}_0)+u^2_{+\alpha}(-{\bf k}_0)+2u_{+\alpha}({\bf k}_0)u_{+\alpha}(-{\bf k}_0)\cos(2 k_0 x+\vartheta_{LR})\bigg]|\varphi_2(z)|^2 \\
&+2\bigg[ 2u_{-\alpha}({\bf k}_0)u_{+\alpha}({\bf k}_0)\cos (k_Tx -\vartheta) -u^2_{-\alpha}({\bf k}_0)\cos \big[(2k_0-k_T)x +(\vartheta_{LR}+\vartheta)\big] \\
&-u^2_{+\alpha}({\bf k}_0)\cos \big[ (2k_0+k_T)x +(\vartheta_{LR}-\vartheta)\big]\bigg]\varphi_1(z)\varphi_2(z) 
\Bigg\},
\end{split}
\end{equation}
This expression can be further simplified by combining the properties  
$u_{-\alpha} (-{\bf k}_0) = - u_{+\alpha} ({\bf k}_0)$ and 
$u_{+\alpha} (-{\bf k}_0) = - u_{-\alpha} ({\bf k}_0)$ with $u^2_{+\alpha} ({\bf k}_\perp)+u^2_{-\alpha} ({\bf k}_\perp)=1$
exhibited in Eq.~(\ref{eqn:u-coefficients-supplementary-material}),
\begin{equation}
\begin{split}
n_C ({\bf r})&=
2 \vert {\cal B}_{\rm dw}\vert^2
\bigg[1-2u_{-\alpha}({\bf k}_0)u_{+\alpha}({\bf k}_0)\cos(2 k_0 x+\vartheta_{LR})\bigg](|\varphi_1(z)|^2+\varphi_2(z)|^2) \\
&+4\vert {\cal B}_{\rm dw}\vert^2\bigg[ 2u_{-\alpha}({\bf k}_0)u_{+\alpha}({\bf k}_0)\cos (k_Tx -\vartheta) -u^2_{-\alpha}({\bf k}_0)\cos \big[(2k_0-k_T)x +(\vartheta_{LR}+\vartheta)\big] \\
&-u^2_{+\alpha}({\bf k}_0)\cos \big[ (2k_0+k_T)x +(\vartheta_{LR}-\vartheta)\big]\bigg]\varphi_1(z)\varphi_2(z) ,
\end{split}
\end{equation}
the normalization requirement
$
N_C
=
\int \mathrm{d}^3 r  
n_C ({\bf r})
$
leads to the normalization constant
\begin{equation}
{\cal B}_{\rm dw} 
=
\sqrt{\frac{N_C}{4L^2_\perp I}},
\end{equation}
where the integral $I$ depends explicitly on the length of the 
system along the $x$ direction, specifically, 
\begin{equation}
I
=
1-\frac{2u_{+\alpha}({\bf k}_0)u_{-\alpha}({\bf k}_0)}{k_0 L_\perp}\sin(k_0 L_\perp)\cos\vartheta_{LR}.
\end{equation}

In the limit that $L_\perp \to \infty$, the integral $I$ tends to one $(I \to 1)$, 
since the functions $\sin (k_0 L_\perp)$ and $\cos \vartheta_{LR}$ are bounded, that is, 
$\vert \sin (k_0 L_\perp) \vert \le 1$ and $\vert \cos\vartheta _{LR}\vert \le 1$. In compact form, the condensate density 
becomes 
\begin{equation}
\begin{split}
\label{eqn:double-well-condensate-supplementary-material}
n_C({\bf r})
&=\frac{N_C}{VI}
\bigg[
1+
2\widetilde \Omega
\cos (2k_0x + \vartheta_{LR} )\bigg]
\bigg[ \cos^2 \bigg(\frac{\pi}{L_z}z\bigg) 
+ 
\sin^2 \bigg(\frac{2\pi}{L_z}z\bigg) \bigg] - \frac{2N_C}{VI}\bigg[ 2\widetilde\Omega \cos(k_Tx-\vartheta) \\
&+\bigg(\frac{1}{2}+\widetilde k_0\bigg) \cos \big[(2k_0-k_T)x +(\vartheta_{LR}+\vartheta)\big] + \bigg(\frac{1}{2}-\widetilde k_0\bigg) \cos \big[(2k_0+k_T)x +(\vartheta_{LR}-\vartheta)\big] \bigg] \cos \bigg(\frac{\pi}{L_z}z\bigg)\sin \bigg(\frac{2\pi}{L_z}z\bigg),
\end{split}
\end{equation}
leading to the dimensionless form, with $V=L^2_\perp L_z$, ${\widetilde n}_C ({\bf r}) = n_C ({\bf r})/n_c$, $k_0\rightarrow {\widetilde k}_0$, $x\rightarrow {\widetilde x}$ and the modifications 
$u_{\pm\alpha} ({\bf k}_0) = u_{\pm\alpha} ({\widetilde k}_0)$, since
these coefficients are dimensionless, as 
shown in Eq.~(\ref{eqn:u-coefficients-supplementary-material}),
and depend only on the $x$ component of momentum. 

A final expression for the dimensionless condensate density
${\widetilde n}_C ({\bf r}) = n_C ({\bf r})/n_c$ in terms of
the condensate fraction $\sigma = N_C/N$ as 
\begin{equation}
\begin{split}
{\widetilde n}_C({\bf r})
&= 
\frac{\sigma}{I}
\bigg[
1+
2\widetilde \Omega
\cos (2k_0x + \vartheta_{LR} )\bigg]
\bigg[ \cos^2 \bigg(\frac{\pi}{L_z}z\bigg) 
+ 
\sin^2 \bigg(\frac{2\pi}{L_z}z\bigg) \bigg] - \frac{2\sigma}{I}\bigg[ 2\widetilde\Omega \cos(k_Tx-\vartheta) \\
&+\bigg(\frac{1}{2}+\widetilde k_0\bigg) \cos \big[(2k_0-k_T)x +(\vartheta_{LR}+\vartheta)\big] + \bigg(\frac{1}{2}-\widetilde k_0\bigg) \cos \big[(2k_0+k_T)x +(\vartheta_{LR}-\vartheta)\big] \bigg] \cos \bigg(\frac{\pi}{L_z}z\bigg)\sin \bigg(\frac{2\pi}{L_z}z\bigg).
\end{split}
\end{equation}

Substitute the wavefunction in Eq.~(\ref{eqn:easy-plane-double-well-wavefunction-supplementary-material}) into the full effective Hamiltonian 
and minimize the energy with respect to $\vartheta$ and $\vartheta_{LR}$, we can fix the random relative phase $\vartheta$ and $\vartheta_{LR}$ in the density profile. In this case, both the independent particle Hamiltonian and the interaction Hamiltonian depend on the relative phase $\vartheta$ and $\vartheta_{LR}$. Then the diagonal term of the independent particle Hamiltonian is,
\begin{equation}
\begin{split}
E_d(\vartheta,\vartheta_{LR})&=\frac{N_C}{2I(\vartheta_{LR})}\bigg[\frac{\hbar^2}{2m}\bigg( \bigg( k_0-\frac{k_T}{2}\bigg)^2 u^2_-({\bf k}_0) + \bigg(k_0+\frac{k_T}{2}\bigg)^2u^2_+({\bf k}_0) - \frac{u_+({\bf k}_0)u_-({\bf k}_0)}{k_0 L_\perp}\bigg(k_0+\frac{k_T}{2}\bigg)^2\sin k_0 L_\perp \cos \vartheta_{LR} \\
&-  \frac{u_+({\bf k}_0)u_-({\bf k}_0)}{k_0 L_\perp}\bigg(k_0-\frac{k_T}{2}\bigg)^2\sin k_0 L_\perp \cos \vartheta_{LR}\bigg)+q\bigg( 1-\frac{2u_+({\bf k}_0)u_-({\bf k}_0)}{k_0 L_\perp}\bigg)\sin k_0 L_\perp \cos \vartheta_{LR}\bigg] \\
&=\frac{N_C}{2I(\vartheta_{LR})}\bigg[\frac{\hbar^2k^2_T}{2m}\bigg( \bigg( \frac{1}{4} - {\widetilde k}^2_0\bigg) + \frac{\widetilde\Omega}{k_0 L_\perp}\bigg(2{\widetilde k}^2_0+\frac{1}{2}\bigg)\sin k_0 L_\perp \cos \vartheta_{LR} \bigg)+q\bigg( 1+\frac{2\widetilde\Omega}{k_0 L_\perp}\bigg)\sin k_0 L_\perp \cos \vartheta_{LR}\bigg]
\end{split}
\end{equation}
and the off-diagonal term is,
\begin{equation}
\begin{split}
E_o(\vartheta,\vartheta_{LR})&=\frac{N_C\Omega}{I(\vartheta_{LR})}\bigg[ u_+({\bf k}_0)u_-({\bf k}_0) \cos \vartheta - \frac{u^2_+({\bf k}_0)}{2k_0 L_\perp}\sin k_0 L_\perp \cos(\vartheta_{LR}-\vartheta)-\frac{u^2_-({\bf k}_0)}{2k_0 L_\perp}\sin k_0 L_\perp \cos(\vartheta_{LR}+\vartheta)\bigg] \\
&=\frac{N_C\Omega k_0L_\perp}{k_0L_\perp+2{\widetilde\Omega}\sin k_0L_\perp\cos \vartheta_{LR}}\bigg[ -\widetilde\Omega \cos \vartheta - \frac{1-2{\widetilde k}_0}{4k_0 L_\perp}\sin k_0 L_\perp \cos(\vartheta_{LR}-\vartheta)-\frac{1+2{\widetilde k}_0}{4k_0 L_\perp}\sin k_0 L_\perp \cos(\vartheta_{LR}+\vartheta)\bigg] \\
\end{split}
\end{equation}
Then the interaction energy becomes
\begin{equation}
\begin{split}
E_I (\vartheta,\vartheta_{LR})&= \frac{c_0N^2_C}{2V^2 I^2} \int \mathrm{d}^3 r  
\bigg\{ \bigg[ 1+2\widetilde \Omega \cos (2k_0x+\vartheta_{LR})\bigg] \bigg[ \cos^2 \bigg(\frac{\pi}{L_z}z\bigg) 
+ 
\sin^2 \bigg(\frac{2\pi}{L_z}z\bigg) \bigg] - 2 \cos \bigg(\frac{\pi}{L_z}z\bigg)\sin \bigg(\frac{2\pi}{L_z}z\bigg) \\
&\times\bigg[ 2\widetilde\Omega \cos(k_Tx-\vartheta)
+\bigg(\frac{1}{2}+\widetilde k_0\bigg) \cos \big[(2k_0-k_T)x +(\vartheta_{LR}+\vartheta)\big] + \bigg(\frac{1}{2}-\widetilde k_0\bigg) \cos \big[(2k_0+k_T)x +(\vartheta_{LR}-\vartheta)\big] \bigg]\bigg\}^2
\end{split}
\end{equation}
and total energy becomes
\begin{equation}
E(\vartheta,\vartheta_{LR}) = E_d(\vartheta,\vartheta_{LR}) + E_o(\vartheta,\vartheta_{LR}) + E_I(\vartheta,\vartheta_{LR})
\end{equation}
The relative phase $\vartheta$, $\vartheta_{LR}$ were determined by minimizing the free energy $E(\vartheta,\vartheta_{LR})$ numerically, resulting in $\vartheta=0$. The energy functional contains a rapid oscillation at the underlying period $\lambda_T$ as the system size $L_\perp$ is varied. In the Fig. 4 of the main text, $\vartheta_{LR}$ equals to $0$ when the total energy is minimized with $k_T L_\perp = 250$. $\vartheta_{LR} = \pi$ achieved similar results for some other value of $k_T L_\perp$. 

Since the density profile along $z$ direction is only originated from the box trap, we can integrate out $z$ direction and scale the density to a 2-dimensional condensate density $n_\perp= N/L^2_\perp$,
\begin{equation}
{\widetilde n}_C({\bf r}_\perp) = \frac{\sigma}{I}\bigg[ 1+2\widetilde \Omega \cos (2k_0x+\vartheta_{LR}) \bigg], 
\end{equation}
where ${\widetilde n}_C({\bf r}_\perp)=\int {\widetilde n}_C({\bf r})  \mathrm{d}z /n_\perp$.
This density profile describes an easy-plane nematic density wave 
with period $\lambda= \pi/k_0$. 
Again, for fixed values of the interaction parameters $c_0$ and $c_2$, 
$\sigma$ is a function of ${\widetilde q}$ and 
${\widetilde \Omega}$. The condensate fraction $\sigma$ for 
the easy-plane nematic double-well phase tends to zero when the phase
boundary ${\widetilde q}_{c} ({\widetilde \Omega})$ is approached
(see Fig.~2 of the main text), since for 
$ {\widetilde q} > {\widetilde q}_{c} ({\widetilde \Omega})$ the
easy-axis nematic phase takes over. However, we use the value of 
$\sigma = 0.7$ for ${\widetilde \Omega} = 1/4$ in the double-well case, 
because we discussed here only an approximate real space Hamiltonian for easy-plane
nematic phases far below the phase boundary 
${\widetilde q}_{c} ({\widetilde \Omega})$, where the condensate 
fraction is closer to one. We choose the same condensate fraction $(\sigma = 0.7)$ and position along $z$ $(z = L/16)$
to plot the local condensate densities of the single-well and double-well phases
in Fig.~4 of the main text, since this facilitates a comparison
of the changes that occur in the amplitude and periods between the single-well and double-well nematic phases.



\begin{thebibliography}{99}
 
%
\bibitem{ueda-2013}
D. M. Stamper-Kurn and M. Ueda, 
Spinor Bose Gases: Symmetries, Magnetism, and Quantum dynamics, 
Rev. Mod. Phys. {\bf 85}, 1191 (2013).
%
	
%
\bibitem{spielman-2009} 
I. B. Spielman,
Raman Processes and Effective Gauge Potentials
Phys. Rev. A {\bf 79}, 063613 (2009).
%

%
\bibitem{dalibard-2010}
G. Juzeli${\bar {\rm u}}$nas, J. Ruseckas, and J. Dalibard,
Generalized Rashba-Dresselhaus Spin-Orbit Coupling for Cold Atoms
Phys. Rev. A {\bf 81}, 053403 (2010).
%


%
\bibitem{spielman-2011}
Y. J. Lin, K. Jim{\'e}nez-Garc{\'i}a, and I. B. Spielman, 
Spin-Orbit-Coupled Bose-Einstein Condensates, 
Nature (London) {\bf 471}, 83 (2011).
%

%
\bibitem{sademelo-2011}
M. Chapman and C. S\'a de Melo, 
Atoms Playing Dress-Up,
Nature (London) {\bf 471}, 41 (2011).
%

%
\bibitem{pan-2014}
S.-C. Ji, J.-Y Zhang, L. Zhang, Z.-D. Du, W. Zheng, Y.-J. Deng, H. Zhai,
S. Chen, and J.-W. Pan,
Experimental Determination of the Finite-Temperature
Phase Diagram of a Spin-Orbit-Coupled Bose gas,
Nature Physics {\bf 10}, 314 (2014) 
%

%
\bibitem{ketterle-2017}
J.-R. Li, J. Lee, W. Huang, S. Burchesky, B. Shteytnas, F. C. Top, 
A. O. Jamison, and W. Ketterle,
A Stripe Phase with Supersolid Properties in Spin-Orbit-Coupled
Bose-Einstein Condensates,
Nature {\bf 543}, 91 (2017).
%

%
\bibitem{zhai-2015}
H. Zhai,
Degenerate Quantum Gases with Spin-orbit Coupling: a Review,
Rep. Prog. Phys. {\bf 78}, 026001 (2015).
%

%
\bibitem{demarco-2015}
M. DeMarco and H. Pu,
Angular Spin-orbit Coupling in Cold Atoms,
Phys. Rev. A {\bf 91}, 033630 (2015).
%

%
\bibitem{zhang-2019}
D. Zhang, T. Gao, P. Zou, L. Kong, R. Li, X. Shen, X. Chen, S. Peng, M. Zhan, H. Pu, and K. Jiang,
Ground-State Phase Diagram of a Spin-orbital-angular-momentum Coupled Bose-Einstein Condensate,
Phys. Rev. Lett. {\bf 122}, 110402 (2019).
%

%
\bibitem{ye-2017}
S. Kolkowitz, S. L. Bromley, T. Bothwell, M. L. Wall, G. E. Marti, A. P. Koller, X. Zhang, A. M. Rey, and J. Ye,
Spin-orbit-coupled Fermions in an Optical Lattice Clock,
Nature {\bf 542}, 66 (2017).
%

%
\bibitem{ye-2018}
S. L. Bromley, S. Kolkowitz, T. Bothwell, D. Kedar, A. Safavi-Naini, M. L. Wall, C. Salomon, A. M. Rey, and J. Ye,
Dynamics of Interacting Fermions under Spin-orbit Coupling in an Optical Lattice Clock,
Nature Physics {\bf 14}, 399 (2018).
%

%
\bibitem{campbell-2016}
D. L. Campbell, R. M. Price, A. Putra, A. Vald\'es-Curiel, D. Trypogeorgos, and I. B. Spielman,
Magnetic Phases of Spin-1 Spin-orbit-coupled Bose Gases,
Nature Communications {\bf 7}, 10897 (2016).
%

%
\bibitem{ketterle-1998}
J. Stenger, S. Inouye, D. M. Stamper-Kurn, H.-J. Miesner, A. P. Chikkatur, 
and W. Ketterle, 
Spin Domains in Ground-State Bose-Einstein Condensates, 
Nature (London) {\bf 396}, 345 (1998).
%

%
\bibitem{machida-1998}
T. Ohmi and K. Machida, 
Bose-Einstein Condensation with Internal Degrees of Freedom in Alkali 
Atom Gases, 
J. Phys. Soc. Jpn {\bf 67}, 1822 (1998).
%

%
\bibitem{zhou-2004}
M. Snoek and F. Zhou,
Microscopic Wave Functions of Spin-Singlet and Nematic Mott States 
of Spin-One Bosons in High-Dimensional Bipartite Lattices, 
Phys. Rev. B {\bf 69}, 094410 (2004).
%
%
\bibitem{demler-2003}
A. Imambekov, M. D. Lukin, and E. Demler, 
Spin-Exchange Interactions of Spin-One Bosons in Optical Lattices: 
Singlet, Nematic, and Dimerized Phases, 
Phys. Rev. A {\bf 68}, 063602 (2003).
%

%
\bibitem{affleck-2004}
F. Zhou, M. Snoek, J. Wiemer, and I. Affleck, 
Magnetically Stabilized Nematic Order: 
Three-Dimensional Bipartite Optical Lattices, 
Phys. Rev. B {\bf 70}, 184434 (2004).
%

%
\bibitem{lett-2007}
A. T. Black, E. Gomez, L. D. Turner, S. Jung, and P. D. Lett,
Spinor Dynamics in an Antiferromagnetic Spin-1 Condensate,
Phys. Rev. Lett. {\bf 99}, 070403 (2007).
%

%
\bibitem{lett-2009}
Y. Liu, S. Jung, S. E. Maxwell, L. D. Turner, E. Tiesinga, and P. D. Lett, 
Quantum Phase Transitions and Continuous Observation of Spinor Dynamics 
in an Antiferromagnetic Condensate, 
Phys. Rev. Lett. {\bf 102}, 125301 (2009).
%

%
\bibitem{raman-2011}
E. M. Bookjans, A. Vinit, and C. Raman,
Quantum Phase Transition in an Antiferromagnetic Spinor Bose-Einstein Condensate,
Phys. Rev. Lett. {\bf 107}, 195306 (2011).
%

%
\bibitem{gerbier-2012}
D. Jacob, L. Shao, V. Corre, T. Zibold, L. De Sarlo, E.
Mimoun, J. Dalibard, and F. Gerbier, 
Phase Diagram of Spin-1 Antiferromagnetic Bose-Einstein Condensates, 
Phys. Rev. A {\bf 86}, 061601 (2012).
%

%
\bibitem{gerbier-2016}
T. Zibold, V. Corre, C. Frapolli, A. Invernizzi, J. Dalibard, and F. Gerbier,
Spin-Nematic Order in Antiferromagnetic Spinor Condensates,
Phys. Rev. A {\bf 93}, 023614 (2016).
%

%
\bibitem{borgh-2014}
M. O. Borgh, J. Lovegrove, and J. Ruostekoski,
Imprinting a Topological Interface Using Zeeman Shifts in a Atomic Spinor Bose-Einstein Condensate,
New J. Phys. {\bf 16}, 053046 (2014).
%

%
\bibitem{symes-2017}
L. M. Symes and P. B. Blakie,
Nematic Ordering Dynamics of an Antiferromagnetic Spin-1 Condensate,
Phys. Rev. A {\bf 96}, 013602 (2017).
%

%
\bibitem{kang-2019}
S. Kang, S. W. Seo, H. Takeuchi, and Y. Shin,
Observation of Wall-Vortex Composite Defects in a Spinor Bose-Einstein Condensate,
Phys. Rev. Lett. {\bf 122}, 095301 (2019).
%

%
\bibitem{andreev-1984}
A. F. Andreev and I. A. Grishchuk, 
Spin Nematics, 
Zh. Eksp. Teor. Fiz. {\bf 87}, 467 (1984) 
[Sov. Phys. JETP {\bf 60}, 267 (1984)].
%

%
\bibitem{vinit-2013}
A. Vinit, E. M. Bookjans, C. A. R. S\'a de Melo, and C. Raman,
Antiferromagnetic Spatial Ordering in a Quenched One-Dimensional Spinor Gas,
Phys. Rev. Lett. {\bf 110}, 165301 (2013).
%

%
\bibitem{vinit-2018}
A. Vinit and C. Raman,
Hanbury Brown–Twiss correlations and multi-mode dynamics in quenched, inhomogeneous density spinor Bose–Einstein condensates,
New J. Phys. {\bf 20}, 095003 (2018).
%

%
\bibitem{sadler-2006}
L. E. Sadler, J. M. Higbie, S. R. Leslie, M. Vengalattore, and D. M. Stamper-Kurn.
Spontaneous Symmetry Breaking in a Quenched Ferromagnetic Spinor Bose-Einstein Condensate,
Nature {\bf 443}, 312 (2006).
%

%
\bibitem{lucke-2011}
B. L{\"{u}}cke, M. Scherer, J. Kruse, L. Pezz{\'{e}}, F. Deuretzbacher, P. Hyllus, O. Topic, J. Peise, W. Ertmer, J. Arlt, L. Santos, A. Smerzi, and C. Klempt,
Twin Matter Waves for Interferometry Beyond the Classical Limit,
Science {\bf 334}, 773 (2011).
%

%
\bibitem{gross-2011}
C. Gross, H. Strobel, E. Nicklas, T. Zibold, N. Bar-Gill, G. Kurizki, and M. K. Oberthaler,
Atomic Homodyne Detection of Continuous-variable Entangled Twin-atom States,
Nature {\bf 480}, 219 (2011)
%

\bibitem{bookjans-2011}
E. M. Bookjans, C. D. Hamley, and M. S. Chapman,
Strong Quantum Spin Correlations Observed in Atomic Spin Mixing,
Phys. Rev. Lett. {\bf 107}, 210406 (2011).
%

%
\bibitem{galitski-2008}
T. D. Stanescu, B. Anderson, and Victor Galitski,
Spin-Orbit-Coupled Bose-Einstein Condensates
Phys. Rev. A {\bf 78}, 023616 (2008).
%


%
\bibitem{ho-2011}
T.-L. Ho and S. Zhang,
Bose-Einstein Condensates with Spin-Orbit Interaction
Phys. Rev. Lett. {\bf 107}, 150403 (2011).
%

%
\bibitem{stringari-2012}
Y. Li, L. P. Pitaevskii, and Sandro Stringari,
Quantum Tricriticality and Phase Transitions in Spin-Orbit-Coupled 
Bose-Einstein Condensates,
Phys. Rev. Lett. {\bf 108}, 225301 (2012).
%

%
\bibitem{baym-2012}
T. Ozawa, and G. Baym,
Stability of Ultracold Atomic Bose Condensates with
Rashba Spin-Orbit Coupling against Quantum and Thermal Fluctuations,
Phys. Rev. Lett. {\bf 109}, 025301 (2012).
%

%
\bibitem{stringari-2013}
Y. Li, G. I. Martone, L. P. Pitaevskii, and S. Stringari,
Superstripes and the Excitation Spectrum of a Spin-orbit-coupled 
Bose-Einstein Condensate,
Phys. Rev. Lett. {\bf 110}, 235302 (2013).
%

%
\bibitem{yamamoto-2017}
D. Yamamoto, I. B. Spielman, and C. A. R. S\'a de Melo,
Quantum Phases of Two-component Bosons with Spin-orbit Coupling in Optical Lattices,
Phys. Rev. A {\bf 96}, 061603(R) (2017).
%

%
\bibitem{chaikin-1995}
P. M. Chaikin and T. C. Lubensky,
{\it Principles of condensed matter physics},
(Cambridge University Press, UK, 1995).
%

%
\bibitem{treutlein-2009}
P. B\"ohi, M. F. Riedel, J. Hoffrogge, J, Reichel, T. W. H\"ansch and P. Treutlein,
Coherent Manipulation of Bose-Einstein Condensates with State-Dependent
Microwave Potentials on an Atom Chip,
Nature Physics {\bf 5}, 592 (2009).
%

%
\bibitem{supplementary-material}
More details can be found in the supplementary material, which includes Refs. [41] and [47].
%

%
\bibitem{ho-1998}
T.-L. Ho, 
Spinor Bose Condensates in Optical Traps,
Phys. Rev. Lett. {\bf 81}, 742 (1998).
%

%
\bibitem{seo-2015}
S. W. Seo, S. Kang, W. J. Kwon, and Y. Shin, 
Half-Quantum Vortices in an Antiferromagnetic Spinor Bose-Einstein Condensate,
Phys. Rev. Lett. {\bf 115}, 015301 (2015).
%

%
\bibitem{lagoudakis-2009}
K. G. Lagoudakis, T. Ostatnick\'y, A. V. Kavokin, Y. G. Rubo, R. Andr\'e and B. Deveaud-Pl\'edran,
Observation of Half-Quantum Vortices in an Exciton-Polariton Condensate,
Science 326 (5955), 974 (2009).
%

%
\bibitem{maeno-2011}
J. Jang, D. G. Ferguson, V. Vakaryuk, R. Budakian, S. B. Chung, P. M. Goldbart and Y. Maeno,
Observation of Half-Height Magnetization Steps in Sr$_2$RuO$_4$,
Science 331 (6014), 186 (2011).
%

%
\bibitem{bloch-2006}
F. Gerbier, A. Widera, S. F\"olling, O. Mandel, and I. Bloch,
Resonant Control of Spin Dynamics in Ultracold Quantum Gases by Microwave Dressing,
Phys. Rev. A {\bf 73}, 041602(R) (2006).
%
\end{thebibliography}

\begin{thebibliography}{99}
%
\bibitem{treutlein-2009}
P. B\"ohi, M. F. Riedel, J. Hoffrogge, J, Reichel, T. W. H\"ansch and P. Treutlein,
Coherent Manipulation of Bose-Einstein Condensates with State-Dependent
Microwave Potentials on an Atom Chip,
Nature Physics {\bf 5}, 592 (2009).
%

%
\bibitem{bloch-2006}
F. Gerbier, A. Widera, S. F\"olling, O. Mandel, and I. Bloch,
Resonant Control of Spin Dynamics in Ultracold Quantum Gases by Microwave Dressing,
Phys. Rev. A {\bf 73}, 041602(R) (2006).
%

\end{thebibliography}
\end{document}